\documentclass[]{revtex4}
\usepackage[]{graphicx}
\usepackage[]{graphics}
\usepackage{latexsym}
\usepackage{amssymb}

\def\lsim{\lower.5ex\hbox{$\; \buildrel < \over \sim \;$}}
\def\gsim{\lower.5ex\hbox{$\; \buildrel > \over \sim \;$}}
\def\ch{\lower-0.55ex\hbox{--}\kern-0.55em{\lower0.15ex\hbox{$h$}}}
\def\lh{\lower-0.55ex\hbox{--}\kern-0.55em{\lower0.15ex\hbox{$\lambda$}}}
\def\lsim{\lower.5ex\hbox{$\; \buildrel < \over \sim \;$}}
\def\gsim{\lower.5ex\hbox{$\; \buildrel > \over \sim \;$}}
\def\AHR{analogue Hawking radiation}

\def\egam{$\left[{\cal E},{\gamma}\right]$}
\def\pth{\left[{\cal E},\lambda,\gamma\right]}
\def\ptw{\left[{\cal E},\lambda\right]}

\def\pthp{\left({\cal E},\lambda,\gamma\right)}
\def\ptwp{\left({\cal E},\lambda\right)}

\begin{document}

\title{Black-Hole Accretion Disc as an Analogue Gravity Model}
\date{\today}

\author{Tapas Kumar Das}
\email{tapas@mri.ernet.in}
\homepage{http://www.mri.ernet.in/~tapas}
\affiliation{Harish Chandra Research Institute, Allahabad-211 019, India.}

\author{Neven Bili\'c}
\email{bilic@thphys.irb.hr}
\homepage{http://thphys.irb.hr/Djelat/bilic.htm}
\affiliation{Rudjer Bo\v{s}kovi\'{c} Institute, 10002 Zagreb, Croatia.}

\author{Surajit Dasgupta}
\email{surajit@tifr.res.in}
\homepage{http://www.tifr.res.in/~surajit}
\affiliation{Tata Institute of Fundamental Research, Mumbai 400 005, India.}

\begin{abstract}
\noindent
We formulate and solve the equations governing the transonic behaviour 
of a general relativistic black-hole accretion disc with non-zero advection 
velocity. We demonstrate that a relativistic Rankine-Hugoniot shock may 
form  leading to the formation of accretion powered outflow. We show 
that the critical points of transonic discs generally do not coincide 
with the corresponding sonic points. The collection of such sonic points 
forms an axisymmetric hypersurface, generators of which are the acoustic 
null geodesics, i.e. the phonon trajectories. Such a surface is shown 
to be identical with  an acoustic event horizon. The acoustic surface 
gravity and the corresponding analogue horizon temperature $T_{AH}$ at 
the acoustic horizon are then computed in terms of fundamental accretion 
parameters. Physically, the analogue temperature is associated with the 
thermal phonon radiation analogous to the Hawking radiation of the 
black-hole horizon.Thus, an axisymmetric black-hole accretion disc is 
established as a natural example of the classical analogue gravity model, 
for which two kinds of horizon  exist simultaneously. We have shown that 
for some values of astrophysically relevant accretion parameters, the 
analogue temperature exceeds the corresponding Hawking temperature. We 
point out  that acoustic {\it white holes} can also be generated for a 
multi-transonic black-hole accretion with a shock. Such a white hole, 
produced at the shock, is always flanked by two acoustic black holes 
generated at the inner and the outer sonic points. Finally, we discuss 
possible applications of our work to other astrophysical events which may
exhibit  analogue effects.
\end{abstract}

\maketitle

\section{Introduction \label{sec:intro}}
\subsection{Black-hole analogue \label{subsec:bhanal}}
\noindent
In recent years, strong analogies have been established between the physics
of acoustic perturbations in an inhomogeneous dynamical fluid system, and 
some kinematic features of space-time in general relativity. An effective 
metric, referred to as the `acoustic metric', which 
describes the geometry of the manifold in which  acoustic perturbations 
propagate, can be constructed. This effective geometry can capture the properties of curved
space-time in general relativity. Physical models constructed utilizing such 
analogies are called `analogue gravity models' (for details on analogue 
gravity models, see, e.g. the review articles \cite{barcelo,cardoso},
and the book \cite{novello}). 
\par One of the most significant effects of analogue gravity is the 
`classical black-hole analogue'. Classical black-hole analogue effects may be  observed when acoustic perturbations (sound waves) 
propagate through a classical, dissipationless, inhomogeneous transonic 
fluid. Any acoustic perturbation, dragged by a supersonically 
moving fluid, can never escape upstream by penetrating the `sonic surface'. 
Such a sonic surface is a collection of transonic points in space-time, 
and can act as a `trapping' surface for outgoing {\it phonons}. Hence, the 
sonic surface is actually an {\it acoustic horizon}, which resembles 
 a black-hole event horizon in 
many ways and is generated at the transonic 
point in the fluid flow. The acoustic horizon is essentially a null hyper 
surface, generators of which are the {\it acoustic} null geodesics, i.e. 
the phonons. The acoustic horizon emits acoustic radiation with quasi thermal 
phonon spectra, which is analogous to the actual Hawking radiation. The 
temperature of the radiation emitted from the acoustic horizon is referred 
to as the analogue Hawking temperature. 
\par In his pioneering work, Unruh \cite{unruh} showed that the scalar field 
describing the acoustic perturbations in a transonic barotropic irrotational 
fluid, satisfied a Klein-Gordon type differential equation for the massless 
scalar field propagating in  curved space-time with a metric that closely 
resembles the Schwarzschild metric near the horizon. The acoustic propagation 
through a transonic fluid forms an analogue event horizon located at the 
transonic point. Acoustic waves with a quasi thermal spectrum will be emitted 
from the acoustic horizon and the temperature of such acoustic radiation is 
given by \cite{unruh}
\begin{equation}
T_{\rm AH}=\frac{\hbar}{4{\pi}\kappa_{\rm B}}\left[\frac{1}{c_s}\frac{{\partial}{u_{\perp}^2}}{\partial{\eta}}
\right]_{\rm Acoustic~Horizon}\, ,
\label{eq1}
\end{equation}
where
$\kappa_{\rm B}$ is  Boltzmann's constant, ${\hbar}$ is the Planck constant
devided by $2\pi$,
$c_s$  the speed of sound,  $u_{\perp}$  the component of the flow velocity 
normal to the acoustic horizon and $ \partial/\partial {\eta}$ represents 
the normal derivative. The temperature $T_{\rm AH}$ defined by Eq. (\ref{eq1}) 
is the acoustic analogue of the usual Hawking temperature $T_{\rm H}$:
\begin{equation}
T_{\rm H}=\frac{{\hbar}c^3}{8{\pi}\kappa_{\rm B}GM_{\rm bh}}
\label{eq2}
\end{equation}
and hence
$T_{\rm AH}$
is referred to as the {\em analogue  Hawking temperature}. In eq. (\ref{eq2}), 
$M_{bh}$ is the black-hole mass, $G$ Newton's gravitational constant, $c$ the 
velocity of light in the vacuum. Note that the sound speed in  Eq. 
(\ref{eq1}) in Unruh's original treatment  \cite{unruh} was assumed constant 
in space.
\par Unruh's work \cite{unruh} was followed by other important papers 
\cite{jacob,95unruh,visser,99jacob,bilic}. A more general treatment of the 
classical analogue radiation for Newtonian fluid was discussed by Visser 
\cite{visser} who considered a general barotropic, inviscid fluid. The  
acoustic metric for a point sink was shown to be conformally related to the 
Painlev\'e-Gullstrand-Lema\^\i{}tre  form of the Schwarzschild metric and a more general
expression for analogue temperature was obtained, where unlike  Unruh's original
expression \cite{unruh}, the speed of sound  was allowed to depend on space coordinates.
In order to determine the analogue Hawking temperature of a classical analogue 
system, one needs to know the location of the acoustic horizon, the velocity of the 
fluid and the speed of sound and their space gradients at the acoustic horizon.
\par In the analogue gravity systems discussed above, the fluid flow is 
non-relativistic in flat Minkowski space, whereas the sound wave propagating through 
the non-relativistic fluid is coupled to a curved pseudo-Riemannian metric. This 
approach has been extended to relativistic fluids \cite{bilic} by incorporating 
the general relativistic dynamics of the fluid flow. Since the introduction of 
viscosity may destroy  Lorenz invariance, the acoustic analogue is best 
studied in a vorticity free dissipationless fluid.

\subsection{Transonic accretion as a black-hole analogue: the motivation. 
\label{subsec:motivation}}
The process by which any gravitating, massive, astrophysical object captures its 
surrounding fluid is called accretion. If  $c_s(r)$ is the local speed of sound 
and $u(r)$ is the instantaneous radial  velocity  of the accreting fluid, moving 
along a space curve parameterized by $r$,  then the local Mach number of the fluid 
can be defined as $M(r)=u(r)/c_s(r)$. The flow will be locally subsonic or supersonic, according to $M(r) < 1$ or $M(r) >1$, i.e. according to $u(r)<c_s(r)$ or 
$u(r)>c_s(r)$. The flow is transonic if at any moment it crosses $M=1$. This 
happens when a subsonic to supersonic or a supersonic to subsonic transition takes 
place either continuously or discontinuously. The points where such crossing takes 
place continuously are called sonic points, and the points of discontinuous 
transition are called shocks or discontinuities.
 
If the accreting material is assumed to be at rest far from the black hole, the 
flow must exhibit transonic behaviour in order to satisfy the inner boundary
conditions imposed by the event horizon. Since the publication of the seminal paper 
by Bondi in 1952 \cite{bondi}, the transonic behaviour of accreting fluid onto 
compact astrophysical objects has been extensively studied in the astrophysics 
community. Similarly, Unruh's paper \cite{unruh} initiated a substantial number of 
works in the theory of analogue Hawking effects with diverse fields of application
\cite{novello,barcelo}. However, except for Moncrief \cite{monc} \footnote{Moncrief's 
work is of particular importance since it provided the idea of sonic geometry for 
spherical black-hole accretion, well before the discovery of analogue gravity.} 
and Anderson \cite{anderson}, until recently no attempt was  made to bridge the 
astrophysical black-hole accretion and the theory of {\AHR},  by providing 
a self-consistent 
study of {\AHR} for real astrophysical fluid flows, i.e. by establishing the fact 
that accreting black holes can be considered as a natural example of analogue system. 
Since both the theory of transonic astrophysical accretion and the theory of {\AHR}  
are based on general relativity, it is almost self-evident that accreting black holes 
can be considered as a natural example of analogue system. 

 Motivated by the above mentioned arguments, it has recently been shown 
\cite{dascqg,topu} that a spherically accreting astrophysical black-hole system 
is a unique example of classical analogue gravity model which exhibits both the 
black-hole event horizon and  the analog acoustic horizon. Hence, an accreting 
astrophysical black hole may be considered  an ideal candidate to 
study these two different types of horizons theoretically and to compare their properties. 

 Analogue effects in an idealized axisymmetric system have recently been 
investigated  \cite{abraham}. There, the axisymmetric astrophysical accretion 
has been modelled by a rotating fluid disc of constant thickness with the
distribution of matter in the disc independent of the $z$ coordinate along the 
axis of rotation. In this paper, we  study analogue effects in a more realistic 
astrophysical system in which the accretion disc thickness is calculated by 
solving the relativistic Euler equation in the vertical direction. The expression 
for the local disc height obtained in this way is a function of the local fluid 
velocity, the local speed of sound and the space coordinate $r$. As will be 
shown in section \ref{sec:tbhadgr_cpsp}, in our realistic disc model, in contrast to the 
disc with constant thickness, the sonic points  do not coincide with the critical 
points, and the overall picture may also differ from that in \cite{abraham}. 
\par In the following sections, we  describe how to model a general relativistic,
axially symmetric, multi-transonic flow of perfect fluid accreting onto  
astrophysical black holes. We formulate and solve the basic equations in 
section \ref{sec:tbhadgr}, and then explore the transonic behaviour of the flow in section \ref{sec:gcs}. 
In section \ref{sec:sfra} we  show that a relativistic standing shock 
wave can form in the accretion disc if the flow is potentially multi-transonic. 
We discuss   the relevant acoustic geometry in detail in section \ref{sec:agahsg}. In section \ref{sec:atshbhad}, we  show how to calculate the analogue 
Hawking temperature. 
Finally, we conclude the paper with section \ref{sec:discussion}. 

\section{Transonic black-hole accretion disc in general relativity \label{sec:tbhadgr} }
\subsection{Multi-transonic flow \label{subsec:tbhadgr_mtf}} 
For the flow of matter with non-zero angular momentum density, accretion phenomena are
 studied employing axisymmetric configuration.
Accreting matter is thrown into circular orbits around the central accretor,
leading to the formation of  accretion discs. The pioneering contribution to 
study the properties of general relativistic black-hole accretion discs may be 
attributed to two classic papers \cite{bardeen,nt73}. For certain values 
of the intrinsic angular momentum density of accreting material, the number of 
sonic points, unlike in spherical accretion,
 may exceed one, and accretion is called `multi-transonic'. 
The study of  multi-transonic flow was initiated by 
Abramowicz and Zurek \cite{zurek}.
Subsequently, multi-transonic behaviour in black-hole accretion discs have been 
expansively studied
\cite{matsumoto,lu,bozena,fukue,kato,sandy,kafatos,pariev,peitz,luetal,apj1,mnras4}.
Typically, the outermost sonic point lies close 
to the corresponding Bondi radius. The innermost sonic point and the middle sonic 
point exist within and outside the marginally stable orbit, respectively, for the 
general relativistic as well as for the post-Newtonian model of accretion flow.
 The location of the sonic points can be calculated 
as a function of the specific flow 
energy ${\cal E}$ (Bernoulli's constant), the specific angular momentum $\lambda$ 
and the inflow polytropic index $\gamma$. 
The literature on multi-transonic flow  usually 
deals with low angular momentum accretion flow. Sub-Keplerian\footnote{The 
`Keplerian' angular momentum refers  to the value of angular 
momentum of a rotating fluid for which the centrifugal force exactly compensates for the 
gravitational attraction. If the angular momentum distribution is 
sub-Keplerian, accretion flow will possess non-zero advective velocity.}
weakly rotating flows are exhibited in various physical situations, such as 
detached binary systems fed by accretion from the so-called OB stellar winds 
\cite{ilashu,liang},
 semi-detached low-mass non-magnetic binaries \cite{bisikalo}, and super-massive 
black holes fed by accretion from slowly rotating central stellar clusters 
\cite{ila,ho}. Even for a standard Keplerian accretion 
disc, turbulence may produce such low angular momentum flow (see, e.g. \cite{igu}
and references therein).

\subsection{The dynamics \label{sec:tbhadgr_d}}
For  the most general description of fluid flow in strong
gravity, one needs to solve the equations of motion for the
fluid and the Einstein equations. The problem may be 
simplified by assuming the accretion to be non-self-gravitating,
 so that the fluid dynamics may be dealt with in a
metric without back-reactions.
We use the units $G=c=M_{bh}=1$, so that
 radial distances and velocities are scaled in units  
$r_g\equiv\frac{G{M_{bh}}}{c^2}$ and $c$,
respectively, and all other derived quantities are scaled accordingly.

We use the Boyer-Lindquist
coordinates \cite{boyer} with signature $-+++$, and an
azimuthally Lorentz boosted orthonormal tetrad basis corotating
with the accreting fluid. We define $\lambda$ to be the specific
angular momentum of the flow and neglect any gravo-magneto-viscous
non-alignment between $\lambda$ and the black-hole spin angular
momentum.

Let $v_\mu$ be the four velocity of the (perfect) accreting fluid. The energy
momentum tensor $ {T}^{{\mu}{\nu}}$ is then given by
\begin{equation}
{T}_{{\mu}{\nu}}=\left(\epsilon+p\right)v_{\mu}v_{\nu}+pg_{{\mu}{\nu}},
\label{eq3}
\end{equation}
with $\epsilon$ and
$p$ being the fluid energy density and pressure, respectively.

In this paper, 
we  study the inviscid
accretion of hydrodynamic fluid.
Hence, our calculation will be focused on the stationary
axisymmetric solution of the energy momentum 
and baryon number conservation equations
\begin{equation}
{T^{{\mu}{\nu}}}_{;\nu}
=0; 
\;\;\;\;\;
\left({\rho}{v^\mu}\right)_{;\mu}=0,
\label{eq4}
\end{equation}
where $\rho$ is the rest-mass density.
Specifying the metric to be stationary and axially symmetric,
 the two
generators
 $\xi^{\mu}\equiv (\partial/\partial t)^{\mu}$ and 
 $\phi^{\mu}\equiv (\partial/\partial \phi)^{\mu}$ of the temporal and
axial isometry, respectively, are
Killing vectors.

We consider the flow to be
`advective', i.e. to possess considerable radial three-velocity.
The above-mentioned advective velocity, which we hereafter denote by $u$
and  consider it to be confined on the equatorial plane, is essentially the 
three-velocity component perpendicular to the set of hypersurfaces
$\{\Sigma_v\}$ defined by
$v^2=const$, where $v$ is the magnitude of the 3-velocity.
Each $\Sigma_v$ is timelike since 
its normal $\eta_{\mu}\propto \partial_{\mu} v^2$ 
is spacelike and may be normalized as
$\eta^{\mu}\eta_{\mu}=1$. 

We then define the specific angular momentum $\lambda$ and the angular
velocity $\Omega$ as
\begin{equation}
\lambda=-\frac{v_\phi}{v_t}; \;\;\;\;\;
\Omega=\frac{v^\phi}{v^t}
=-\frac{g_{t\phi}+\lambda{g}_{tt}}{{g_{\phi{\phi}}+\lambda{g}_{t{\phi}}}}\, ,
\label{eq5}
\end{equation}

The metric on the equatorial plane is given by \cite{nt73}
\begin{equation}
ds^2=g_{{\mu}{\nu}}dx^{\mu}dx^{\nu}=-\frac{r^2{\Delta}}{A}dt^2
+\frac{A}{r^2}\left(d\phi-\omega{dt}\right)^2
+\frac{r^2}{\Delta}dr^2+dz^2 ,
\label{eq6}
\end{equation}
where $\Delta=r^2-2r+a^2, ~A=r^4+r^2a^2+2ra^2$,
and $\omega=2ar/A$, $a$ being the Kerr parameter related to the black-hole spin. 
The normalization condition $v^\mu{v}_\mu=-1$, together with  
the expressins for  
$\lambda$ and $\Omega$  in Eq. (\ref{eq5}), provides the relationship between the
advective velocity $u$ and the temporal component of the four velocity
\begin{equation}
v_t=
\left[\frac{Ar^2\Delta}
{\left(1-u^2\right)\left\{A^2-4\lambda arA
+\lambda^2r^2\left(4a^2-r^2\Delta\right)\right\}}\right]^{1/2} .
\label{eq7}
\end{equation}
\subsection{Thermodynamics \label{sec:tbhadgr_td}}
In order to solve Eqs. (\ref{eq4}), we need to specify a realistic equation of
state. In this work, we concentrate on polytropic accretion. However, polytropic 
accretion is not the only choice to describe the general relativistic transonic 
black-hole accretion. Equations of state other than the adiabatic one,  such as 
the isothermal equation \cite{kafatos} or the two-temperature plasma \cite{manmoto}, 
have also been used to study the black-hole accretion flow.

We assume the dynamical in-fall time scale to be short compared with any 
dissipation time scale during the accretion process. To describe the fluid,
 we use a 
 polytropic equation of state  of the form
\begin{equation}
p=K{\rho}^\gamma ,
\label{eq8}
\end{equation}
where the polytropic index $\gamma$ equal to the ratio of the two specific
heats $c_p$ and $c_v$ of the accreting material is assumed to be constant throughout the fluid.
A more realistic model of the flow 
would perhaps require  a variable polytropic index having a 
functional dependence on the radial
distance, i.e. $\gamma=\gamma(r)$. However, we  have performed the
calculations for a sufficiently large range of $\gamma$ and we believe
that all astrophysically relevant
polytropic indices are covered.

The constant $K$ in Eq. (\ref{eq8}) may be 
related to the specific entropy of the fluid,
 provided there is no entropy 
generation during the flow. 
If in addition to (\ref{eq8}) the
Clapeyron equation for an ideal gas 
holds
\begin{equation}
p=\frac{\kappa_B}{{\mu}m_p}{\rho}T\, ,
\label{eq9}
\end{equation}
 where $T$ is the locally measured temperature, $\mu$  the mean molecular weight,
$m_H{\sim}m_p$  the mass of the hydrogen atom, then the specific entropy, i.e. the entropy 
per particle, is given by \cite{landau}
\begin{equation}
\sigma=\frac{1}{\gamma -1}\log K+
\frac{\gamma}{\gamma-1}+{\rm constant} ,
\label{eq10}
\end{equation}
where the constant depends on the chemical composition of the 
accreting material. 
Equation (\ref{eq10}) confirms that $K$ in Eq. (\ref{eq8})
is  a measure of the specific entropy of the accreting matter.

The specific enthalpy of the accreting matter can now be defined as
\begin{equation}
h=\frac{\left(p+\epsilon\right)}{\rho}\, ,
\label{eq11}
\end{equation}
where the energy density $\epsilon$ includes the rest-mass density and the internal 
energy and may be written as
\begin{equation}
\epsilon=\rho+\frac{p}{\gamma-1}\, .
\label{eq12}
\end{equation}
The adiabatic speed of sound is defined by
\begin{equation}
c_s^2=\frac{{\partial}p}{{\partial}{\epsilon}}{\Bigg{\vert}}_{\rm 
constant~entropy}\, .
\label{eq13}
\end{equation}
From Eq. (\ref{eq12}) we obtain
\begin{equation}
\frac{\partial{\rho}}{\partial{\epsilon}}
=\left(
\frac{\gamma-1-c_s^2}{\gamma-1}\right) .
\label{eq14}
\end{equation}
Combination of Eq. (\ref{eq13}) and Eq. (\ref{eq8}) gives
\begin{equation}
c_s^2=K{\rho}^{\gamma-1}{\gamma}\frac{\partial{\rho}}{\partial{\epsilon}}\, ,
\label{eq15}
\end{equation}
Using the above relations, one obtains the expression for the specific enthalpy 
\begin{equation}
h=\frac{\gamma-1}{\gamma-1-c_s^2}\, .
\label{eq16}
\end{equation}
The rest-mass density $\rho$, the pressure $p$, the temperature $T$
of the flow and the energy density $\epsilon$ 
may be expressed in terms of the speed of sound  $c_s$ as
\begin{equation}
\rho=K^{-\frac{1}{\gamma-1}}
\left(\frac{\gamma-1}{\gamma}\right)^{\frac{1}{\gamma-1}}
\left(\frac{c_s^2}{\gamma-1-c_s^2}\right)^{\frac{1}{\gamma-1}},
\label{eq17}
\end{equation}
\begin{equation}
p=K^{-\frac{1}{\gamma-1}}
\left(\frac{\gamma-1}{\gamma}\right)^{\frac{\gamma}{\gamma-1}}
\left(\frac{c_s^2}{\gamma-1-c_s^2}\right)^{\frac{\gamma}{\gamma-1}},
\label{eq18}
\end{equation}
\begin{equation}
T=\frac{\kappa_B}{{\mu}m_p}
\left(\frac{\gamma-1}{\gamma}\right)
\left(\frac{c_s^2}{\gamma-1-c_s^2}\right),
\label{eq19}
\end{equation}
\begin{equation}
\epsilon=
K^{-\frac{1}{\gamma-1}}
\left(\frac{\gamma-1}{\gamma}\right)^{\frac{1}{\gamma-1}}
\left(\frac{c_s^2}{\gamma-1-c_s^2}\right)^{\frac{1}{\gamma-1}}
\left[
1+\frac{1}{\gamma}
\left(
\frac{c_s^2}{\gamma-1-c_s^2}
\right)
\right].
\label{eq20}
\end{equation}
\subsection{Disc geometry and conservation equations \label{sec:tbhadgr_dgce}}
\noindent
We assume that
the disc has a radius-dependent local
thickness $H$, and its central plane coincides with
the equatorial plane of the black hole.
It is a standard practice
in accretion disc theory
to
use the vertically integrated
model in
describing the black-hole accretion discs where the equations of motion
apply to the equatorial plane of the black hole, assuming the flow to
be in hydrostatic equilibrium in the transverse direction. We follow the same 
procedure here. 
The flow variables are averaged over the disc height,
 i.e.
a quantity $y$ used in our model is vertically integrated over the disc height and averaged as
$\bar{y}=\int^H_0 dh\, y /H$.
We follow  \cite{discheight} to derive an expression for the disc height $H$ 
in our flow geometry since the relevant equations in \cite{discheight}  
are non-singular on the horizon and can accommodate both the axial and  
a quasi-spherical flow geometry. In the Newtonian framework, the disc height
 in vertical 
equilibrium is obtained from the $z$ component of the non-relativistic Euler 
equation where all the terms involving velocities and the
higher powers of $\left(({z}/{r}\right)$ are neglected. 
In the case of a general relativistic disc, the vertical pressure
gradient in the comoving frame is compensated by the tidal gravitational
field. We then obtain the disc height 
\begin{equation}
H=\sqrt{\frac{2}{\gamma + 1}} r^{2} \left[ \frac{(\gamma - 1)c^{2}_{c}}
{\{\gamma - (1+c^{2}_{s})\} \{ \lambda^{2}v_t^2-a^{2}(v_{t}-1) \}}\right] ^{\frac{1}{2}} ,
\label{eq21}
\end{equation}
which, by making use of
 Eq. (\ref{eq7}), 
may be be expressed in terms of 
the advective velocity $u$. 

 The temporal  component of the energy momentum tensor conservation equation 
 leads to the
 constancy along each streamline of the flow specific energy 
${\cal E}$ (relativistic analogue of Bernoulli's constant) defined as 
\cite{anderson} 
\begin{equation}
{\cal E}=hv_t. 
\end{equation}

From (\ref{eq7}) and (\ref{eq16}) it follows 
\begin{equation}
{\cal E} =
\left[ \frac{(\gamma -1)}{\gamma -(1+c^{2}_{s})} \right]
\sqrt{\left(\frac{1}{1-u^{2}}\right)
\left[ \frac{Ar^{2}\Delta}{A^{2}-4\lambda arA +
\lambda^{2}r^{2}(4a^{2}-r^{2}\Delta)} \right] } \, .
\label{eq22}
\end{equation}
The rest-mass accretion rate ${\dot M}$ is obtained by integrating the relativistic 
continuity equation (\ref{eq4}). One finds
\begin{equation}
{\dot M}=4{\pi}{\Delta}^{\frac{1}{2}}H{\rho}\frac{u}{\sqrt{1-u^2}} \, ,
\label{eq23}
\end{equation}
Here, we adopt the sign convention that a positive $u$ corresponds to
accretion.
The entropy accretion rate ${\dot S}$ is a quasi-constant
multiple of the mass accretion rate:
\begin{equation}
{\dot S}
 = \left( \frac{1}{\gamma} \right)^{\left( \frac{1}{\gamma-1} \right)}
4\pi \Delta^{\frac{1}{2}} c_{s}^{\left( \frac{2}{\gamma - 1}\right) } \frac{u}{\sqrt{1-u^2}}\left[\frac{(\gamma -1)}{\gamma -(1+c^{2}_{s})}
\right] ^{\left( \frac{1}{\gamma -1} \right) } H .
\label{eq24}
\end{equation}
Note that, in the absence of creation or annihilation of matter,
the mass accretion rate is a constant of motion,
whereas the entropy accretion
rate is not. As the expression for ${\dot S}$ contains the quantity 
$K\equiv p/\rho^\gamma$, which measures  the
specific entropy of the flow, the entropy rate ${\dot S}$ remains constant
throughout the flow {\it only if} the entropy per particle
remains locally unchanged.
This latter condition may be violated if the accretion is
accompanied by a shock.
 Thus, ${\dot S}$ is a
constant of motion for shock-free polytropic accretion and
becomes discontinuous (increases) at the shock location,
if a shock forms in the accretion.
One can solve the two conservation equations for ${\cal E}$ and
${\dot S}$ to obtain the complete accretion profile. 
In this paper, we  concentrate on the 
Schwarzschild metric only. A more general solution for 
the Kerr metric is in progress and will be presented elsewhere.
 For $a=0$,  the expressions  
for ${\cal E}$, ${\dot M}$ and ${\dot S}$ are 
\begin{equation}
{\cal E}_{\rm Schwarzschild}=\left[ \frac{(\gamma -1)}{\gamma -(1+c^{2}_{s})}
 \right]r
\sqrt{\frac{r-2}{r^3-\lambda^2\left(r-2\right)}}
\frac{1}{\sqrt{1-u^2}},
\label{eq25}
\end{equation}
\begin{equation}
\dot M _{\rm Schwarzschild}= 
\frac{4{\pi}{\rho}c_sr^{\frac{3}{2}}u}{\lambda}
\sqrt{
\frac{2\left(\gamma-1\right)\left[r^3-\lambda^2\left(r-2\right)\right]}{\gamma\left[\gamma-\left(1+c_s^2\right)\right]}
}\, ,
\label{eq26}
\end{equation}
\begin{equation}
{\dot S}_{\rm Schwarzschild}=
 4\pi \left( \frac{1}{\lambda} \sqrt\frac{2}{\gamma} \right) 
\left[\frac{c_{s}} {\left (1-\frac{c_{s}^2}{\gamma-1}\right )^{\frac{1}{2}}}
 \right]^{\frac{\gamma+1}{\gamma-1}} u r \left[r^4-\lambda^2 r(r-2)
 \right]^\frac{1}{2},
\label{eq27}
\end{equation}
The corresponding disc height
is given by
\begin{equation}
H_{\rm Schwarzschild}=\frac{c_{s}r}{\lambda} \sqrt{\frac{2(\gamma-1)
(1-u^{2})[r^{3}-\lambda^{2}(r-2)]}{\gamma[\gamma-(1+c_{s}^{2})](r-2)}}\, .
\label{eq28}
\end{equation}
\subsection{Velocity gradients and critical points \label{sec:tbhadgr_vgcp}}
\noindent
By taking the logarithmic derivative of both sides of Eq. (\ref{eq27}) we obtain 
the sound speed gradient as
\begin{equation}
\frac{dc_{s}}{dr}=-\frac{c_{s}(\gamma-1)
\left[\gamma-(1+c^{2}_{s}\right)]}{(\gamma+1)}
\left[ \frac{1}{u} \frac{du}{dr} + {f_{1}}(r,\lambda) \right],
\label{eq29}
\end{equation}
where 
\begin{equation}
{f_{1}}(r,\lambda) = \frac{3r^{3}-2\lambda^{2}r+
3{\lambda^2}}{r^{4}-\lambda^{2}r(r-2)}\, .
\label{eq30}
\end{equation}
Differentiation of both sides of Eq. (\ref{eq25}) and the substitution
 of $(dc_s/dr)$ from Eq. (\ref{eq29}) gives the advective velocity 
gradient 
\begin{equation}
\frac{du}{dr} = \frac{(\frac{2}{\gamma+1})c^{2}_{s} {f_{1} } (r,\lambda)
 - {f_{2} } (r,\lambda)}{\frac{u}{1-u^{2}} - \frac{2c^{2}_{s}}{u(\gamma+1)}}
=\frac{{\cal N}\left(r,\lambda,c_s\right)}{{\cal D}\left(u,c_s\right)}\, ,
\label{eq31}
\end{equation}
where
\begin{equation}
{f_{2} } (r,\lambda) = \frac{2r-3}{r(r-2)} - 
\frac{2r^{3}-\lambda^{2}r+\lambda^{2}}{r^{4}-\lambda^{2}r(r-2)}\, .
\label{eq32}
\end{equation}
A real physical transonic flow must be smooth everywhere,
except possibly at a shock. Hence, if the denominator
${{\cal D}\left(u,c_s\right)}$ of Eq. (\ref{eq31}) 
vanishes at a point, the numerator
${{\cal N}\left(r,\lambda,c_s\right)}$ must also vanish at that point to ensure the
physical continuity of the flow. One therefore arrives at the {\em critical point}
conditions
by  making ${{\cal D}\left(u,c_s\right)}$  and 
${{\cal N}\left(r,\lambda,c_s\right)}$ of Eq. (\ref{eq31})
simultaneously equal to zero.
 We thus obtain the 
critical point conditions as
\begin{equation}
u_c=\pm \sqrt{
\frac
{{f_{2}}(r_c,\lambda)}
{{{f_{1}}(r_c,\lambda)}+{{f_{2}}(r_c,\lambda)}}
}; \;\;\;\;
c_c=\frac{\gamma+1}{2}\left[\frac
{{f_{2}}(r_c,\lambda)}{{f_{1}}(r_c,\lambda)}
\right],
\label{eq33}
\end{equation}
where $u_c\equiv u({r_c})$ and $c_c\equiv c_s (r_c)$, $r_c$ being the
 location 
of the critical point or the so-called `fixed point' of the differential 
equation (\ref{eq31}).
Hereafter, a   
subscript $c$ marks a quantity  
evaluated at $r=r_c$. 
The $+$ or $-$ sign in (\ref{eq33}) corresponds to accretion or
wind, respectively.

Hereafter, we denote by $\pth$ 
our three-parameter space
 characterizing the flow behaviour.
It is understood that ${\ptw}{\subseteq}{\pth}$ 
will be a sub-set of $\pth$ for a 
fixed value of $\gamma$.
A point in the parameter space, with particular 
values of the parameters,
is denoted by  $\pthp$
or $\ptwp$.

The conserved specific energy  can be expressed in terms of
$r_c$
\begin{eqnarray}
{\cal E} & = &  
\left[
\frac
{2{{f_{1}}(r_c,\lambda)}\left(\gamma-1\right)}
{2{{{f_{1}}(r_c,\lambda)}\left(\gamma-1\right)}
+{{{f_{2}}(r_c,\lambda)}\left(\gamma+1\right)}}
\right]
\nonumber \\
& &
\sqrt{
\frac
{r_c^2
\left(r_c-2\right)
\left\{
{{{f_{1}}(r_c,\lambda)}\left(\gamma-1\right)}
+{{{f_{2}}(r_c,\lambda)}\left(\gamma-1\right)}
\right\}
}
{
{{{f_{1}}(r_c,\lambda)}\left(\gamma-1\right)}
\left\{r_c^3-{\lambda^2}\left(r_c-2\right)\right\}
}
} \, ,
\label{eq34}
\end{eqnarray}
For a particular value of $\pthp$, one can now solve Eq. (\ref{eq34}) to find 
 the corresponding  value
of $r_c$.

To determine the behaviour of the solution in the neighbourhood of the critical point,
we need to evaluate the space gradient of 
the advective velocity 
$u'\equiv du/dr$ at the critical point.
Equation (\ref{eq31}) is equivalent to the set of two parametric
 first-order differential equations
\begin{equation}
\frac{du}{d\tau} 
={\cal N}\left(r,\lambda,c_s\right),
\label{eq031}
\end{equation}
\begin{equation}
\frac{dr}{d\tau} = {\cal D}\left(u,c_s\right),
\label{eq131}
\end{equation}
with ${\cal N}$ and ${\cal D}$ vanishing simultaneously at the 
critical  or fixed point.
The value $u'_c$ that  $u'$ takes at the fixed point is obtained 
by applying the L` Hospital's rule to 
Eq. (\ref{eq31}). We obtain a quadratic equation
\begin{equation}
\Psi {u'_c}^2+ {\Re}_1 u'_c+{\Re}_2=0,
\label{eq42}
\end{equation}
with the solutions
\begin{equation}
u'_c=
 \frac{1}{2\Psi} \left[- {\Re}_1 {\pm} \sqrt{{\Re}_1^2{+}
  4{\Re}_2\Psi(u_c,c^{}_{c})} \right],
\label{eq35}
\end{equation}
where
\begin{equation}
\Psi = \left(\frac{\partial {\cal D}}{\partial u} +\frac{\alpha}{u}
\frac{\partial {\cal D}}{\partial c_s}
\right)_c ,
\label{eq36}
\end{equation}
\begin{equation}
{\Re}_1 =
  \left(\alpha
 f_1 \frac{\partial {\cal D}}{\partial c_s}
- \frac{\alpha}{u}
\frac{\partial {\cal N}}{\partial c_s}
\right)_c  ,
\label{eq37}
\end{equation}
\begin{equation}
\Re_2 = -\left(\frac{\partial {\cal N}}{\partial r}+
\alpha f_1 \frac{\partial {\cal N}}{\partial c_s}\right)_c  ,
\label{eq38}
\end{equation}
with
\begin{equation}
\alpha=-\frac{c_{s}(\gamma-1)
\left[\gamma-(1+c^{2}_{s}\right)]}{(\gamma+1)} \, .
\label{eq40}
\end{equation}


\begin{center}
\begin{figure}[h]
\includegraphics[scale=0.6,angle=270.0]{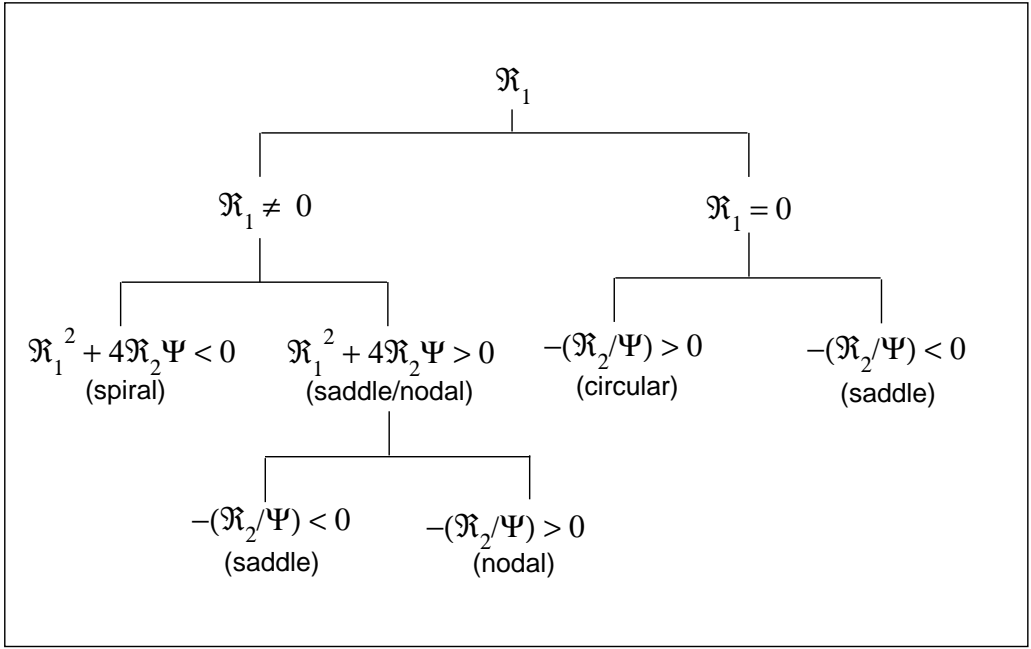}
\caption[]{Classification scheme for critical points. 
For details see section \ref{sec:tbhadgr_vgcp}.}
\label{fig1}
\end{figure}
\end{center}

The solutions are classified according to their
topological behaviour  in the neighbourhood of the
fixed points. 
If $\Re_1$ is non-zero, the inequality ${\Re_1}^2<4\Re_2\Psi$ indicates that the fixed
 point is
of a spiral type, and if  ${\Re_1}^2>4\Re_2\Psi$, the fixed point is  of a saddle type 
for $\Re_2>0$ or a nodal type for $\Re_2<0$. If $\Re_1=0$, there are no
 nodal points, and the spiral points  become of a {\em circular} type when
  $\Re_2<0$.
The saddle-type fixed points in this case  correspond to the condition 
$\Re_2>0$. 
The classification scheme for the critical points is depicted by 
the flow chart diagram shown in Fig. \ref{fig1}.

For a particular value of $\pthp$, Eqs. (\ref{eq25})-(\ref{eq40}) can be solved simultaneously to obtain 
a  transonic solution passing through the critical point represented by
$\pthp$. However, one should note that an `acceptable'  physical 
transonic solution must be globally consistent, i.e. it must connect
 the radial infinity
$r{\rightarrow}\infty$ with the black-hole event horizon $r=2r_g$.
 This  acceptability    constraint 
further demands that  the critical point corresponding to the flow 
should be  
of a saddle or a nodal type. 
This condition is necessary although not sufficient,
 as we discuss in the following sections.
Spiral or centre-type critical points do not 
admit a globally acceptable solution. For further discussion on the classification of critical points, see, eg. \cite{ferrari,kato} and references 
therein.

\subsection{Distinction between critical points and sonic points \label{sec:tbhadgr_cpsp} }
Before we proceed, let us discuss one important issue regarding 
the relation between the critical points and the sonic points in this work.
 From Eq. (\ref{eq33})
  one can calculate the Mach number of the flow at the critical point as
\begin{equation}
M_c=
\sqrt{
\left({\frac{2}{\gamma+1}}\right)
\frac
{{f_{1}}(r_c,\lambda)}
{{{f_{1}}(r_c,\lambda)}+{{f_{2}}(r_c,\lambda)}}
}\, .
\label{eq44}
\end{equation}
Clearly, $M_c$ is generally not equal to one, and for $\gamma\geq 1$, is always less 
than one. Hence  we distinguish a sonic point from a critical point.
In the literature on transonic black-hole accretion discs, the concepts of critical 
and sonic points are often made synonymous by defining an `effective' sound speed 
leading to the `effective'  Mach number (for further details, see, eg.
 \cite{matsumoto,sandy}).
  Such  definitions were proposed as effects of a 
specific disc geometry. We, however, prefer to maintain the usual definition of the Mach 
number for two reasons.
 
First, in the existing literature on transonic disc accretion, 
the Mach number at the critical point  turns out to be a function of 
$\gamma$ only, and hence $M_c$ remains  constant  if $\gamma$ is constant.
 For example, 
using the Paczy\'nski and Wiita \cite{wiita} pseudo-Schwarzschild potential to 
describe the accretion phenomena leads to 
\begin{equation}
M_c=\sqrt{\frac{2}{\gamma+1}}\, .
\label{eq45}
\end{equation}
However, the quantity $M_c$ in Eq. (\ref{eq44}) is clearly a function of $r_c$, and hence, generally, it takes  different 
values for different $r_c$ for multi-transonic accretion, even at 
a fixed value of $\pthp$.
In the following paragraphs we show that the difference between the 
radii of the critical 
point and the sonic point may be quite significant.
 We define the
radial difference as
\begin{equation}
{\Delta}r_c^s=|r_s-r_c|.
\label{eq46}
\end{equation}
The quantity ${\Delta}r_c^s$ may be  
 a complicated  function of $\pthp$, the  form of which can not 
be expressed analytically. 
The radius $r_s$ in Eq. (\ref{eq46}) is the radius of the
sonic point  corresponding to the same $\pthp$ for which the
radius of the critical point $r_c$ is evaluated.
 Note, however, that since $r_s$ is calculated by integrating the 
flow from $r_c$, ${\Delta}r_c^s$ is defined only for saddle-type
critical points. This is because, as we will see in the subsequent sections,
 a physically acceptable transonic solution
can be constructed only through a saddle-type critical point. 
 One can then show that ${\Delta}r_c^s$ can be as large as $10^2$ $r_g$ or even
  more (for details, see section \ref{subsec:gcs_deltar}). 

The second and perhaps the more important reason for keeping $r_c$ and $r_s$ 
distinct
is the following. 
In addition to studying the dynamics of general relativistic transonic 
black-hole accretion, we are also interested in studying the 
analogue Hawking effects for such accretion flow.
We need to identify the 
 location of the acoustic horizon 
 as a radial distance at which  the Mach equals to one, hence, a {\it sonic
  point}, and not a {\em  critical point} 
will be of our particular interest.
 To this end, we first calculate the critical point
  for a particular $\pthp$
following the procedure discussed above, and then we compute the location 
of the sonic point by integrating the flow equations starting from the critical points.
The details of this procedure are provided in the following sections.
Furthermore, the definition of the acoustic metric in terms of 
the sound speed does not seem to be mathematically consistent with the idea of
an `effective' sound speed, irrespective of whether one deals with
 the Newtonian,  post Newtonian, or a relativistic description
of the accretion disc. Hence, we do not adopt
the idea of identifying critical  with  sonic points.
However,  for saddle-type
critical points, $r_c$ and $r_s$ should always have one-to-one correspondence, 
in the sense that
every  critical point that  allows a steady solution to pass through it 
is accompanied by a sonic point, generally at a 
different radial distance
$r$.

It is worth emphasizing that the distinction between critical and  
sonic points is a direct manifestation of the non-trivial 
functional dependence of the disc thickness on
the fluid velocity, the sound speed
and the radial distance.
 In the simplest idealized case when
the disc thickness is assumed to be constant,
one would expect no distinction
between critical and sonic points. 
In this case, as
 has  been demonstrated for a thin disc
 accretion onto the Kerr black hole  \cite{abraham},
  the quantity $\Delta r_c^s$
  vanishes
identically for any astrophysically relevant value of $\cal E$, $\lambda$,
$\gamma$, and 
the Kerr black-hole spin parameter $a$.
\section{Global classification of the $\pth$ space based on the nature of 
the critical points \label{sec:gcs}}
\subsection{Choice of $\left[{\cal E},\gamma\right]$ and the 
classification scheme \label{subsec:gcs_scheme} }
Next, we give a complete classification scheme for the 
critical points and the corresponding flow solutions, for the parameter 
space spanned by all astrophysically relevant values of $\pth$.
We first set the appropriate bounds on {\egam} to model the realistic situations
encountered in astrophysics. 
Since the specific energy ${\cal E}$  includes the rest-mass energy,
${\cal E}=1$ is the lower bound which corresponds to a flow with zero
thermal energy at infinity.
Hence, the values ${\cal E}<1$, corresponding to the negative energy accretion states, would be allowed if a mechanism
for a radiative extraction of the rest-mass energy existed.
The possibility of such an extraction
would in turn imply
viscosity or other dissipative mechanisms in the fluid, 
the properties which would violate  Lorenz invariance.
Since Lorenz invariance is a prerequisite for studying the 
analogue Hawking effects, we  concentrate only on
 non-dissipative flows, 
and hence, we exclude ${\cal E}<1$.
On the other hand, although almost all ${\cal E}>1$ are theoretically allowed,
 large values of ${\cal E}$ represent flows starting from infinity
with very high thermal energy. 
In particular, ${\cal E}>2$ accretion represents enormously
hot flow configurations at very large distance from the black hole,
which are not properly conceivable in realistic astrophysical situations.
Hence, we set $1{\lsim}{\cal E}{\lsim}2$. 

The physical lower bound on the polytropic index is
$\gamma=1$,  which  corresponds to isothermal accretion
where accreting fluid remains optically thin. 
Hence, the values $\gamma<1$ are not realistic in accretion
astrophysics. 
On the other hand,
$\gamma>2$ is possible only for superdense matter
with a very large magnetic
field 
and a direction-dependent anisotropic pressure.
The presence of a magnetic field would in turn
require solving the 
 general relativistic
magneto-hydrodynamic
equations,  which
is beyond the scope of this paper.
Thus, we set 
 $1{\lsim}\gamma{\lsim}2$.  However,  
  astrophysically 
preferred
values of $\gamma$ for realistic black-hole accretion range from 
$4/3$ (ultra-relativistic) 
to $5/3$ (purely non-relativistic flow) \cite{frank}.
 Hence, we mainly focus on the parameter range 
\begin{equation}
\left[1{\lsim}{\cal E}{\lsim}2,~\frac{4}{3}{\le}\gamma{\le}\frac{5}{3}\right].
\label{eq47}
\end{equation}

\begin{figure}
\includegraphics[scale=0.65,angle=-90]{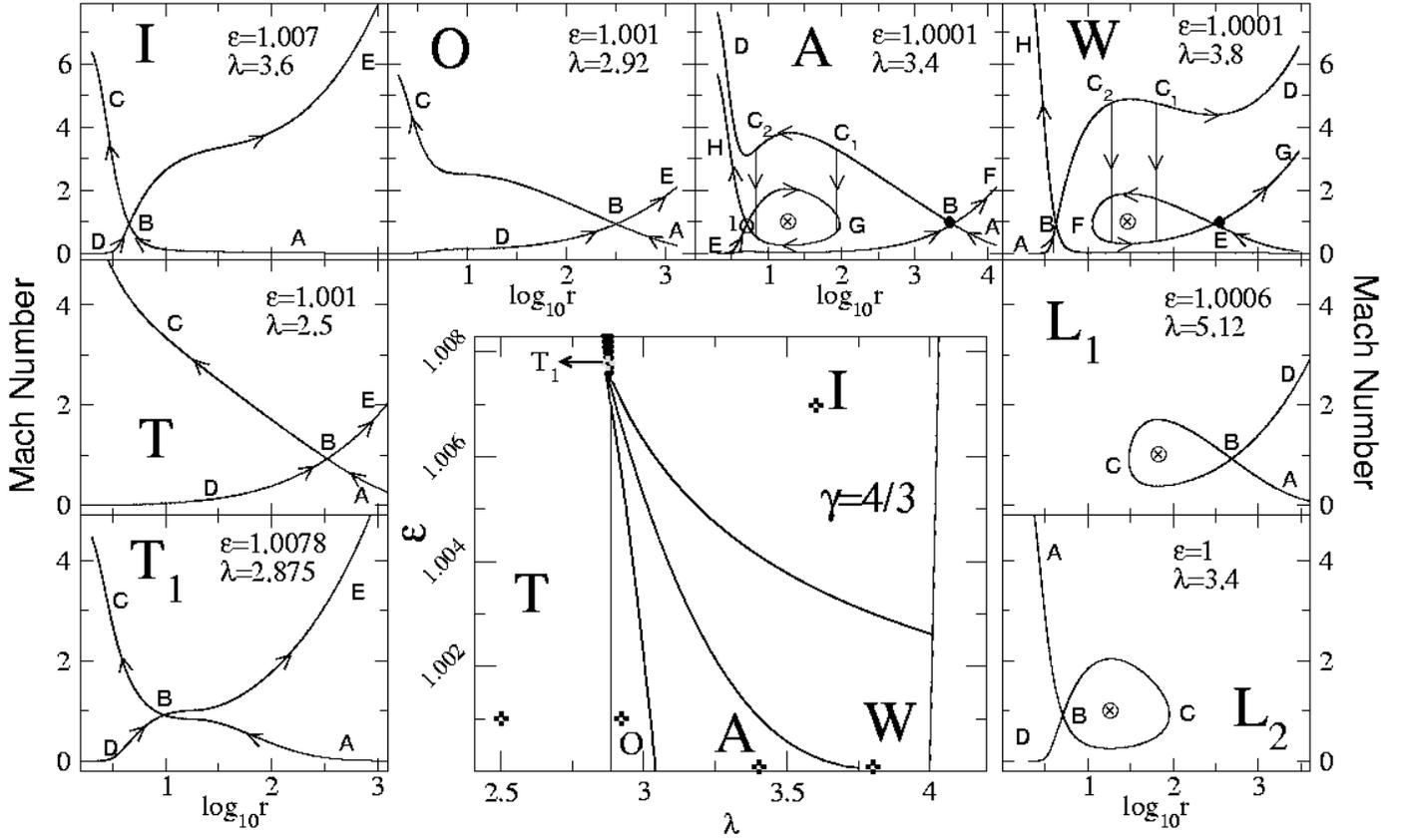}
\caption{Parameter space division for general relativistic accretion and wind.
The side panels show the representative topologies for all distinct critical
 point zones drawn in the central plot.
The bold capitals that mark the side panels correspond to those
that mark the regions of the parameter space in the central plot.
The points in the parameter space for which the side panels are drawn 
are marked in the central plot by  a thick black cross 
with a white square at its centre.
The inward oriented curves marked by arrow heads indicate accretion, 
 while those oriented outwards indicate wind. }
 \label{fig2} 
\end{figure}

Figure \ref{fig2} shows a complete classification scheme for the critical point parameter 
space together with some representative flow topologies. 
The figure has been drawn for the fluid with
$\gamma=4/3$, but a similar figure may be produced 
for any other value of astrophysically relevant $\gamma$.
 The central plot of Fig. \ref{fig2} depicts the classification of $\ptw$
  according 
to the number and nature of the critical points in the flow. 
The side panels show the representative topologies for all distinct critical
 point zones drawn in the central plot.
  The  panel marked by, say, {\bf W}  
 shows a representative topology for the particular $\ptwp$ chosen
from the region marked by {\bf W} in the central plot.
 A thick black cross with a white 
square at its centre in the {\bf W} zone indicates the point in the 
parameter space
$\ptwp$ for which the panel topology {\bf W} is drawn.
The horizontal axis in the  panel figures represents the radial distance
in units of $r_g$
on the logarithmic scale,
whereas the vertical axis represents 
the flow Mach number.  
The parameters ${\cal E}$ and  
$\lambda$ for which  a particular topology is drawn 
are written at the top right corner of the corresponding panel figure.

\subsection{Mono-transonic solutions with one critical point \label{subsec:gcs_mts1cp}}
\noindent
The wedge shaped regions marked by {\bf O} and {\bf I} 
in Fig. \ref{fig2} represent the values
of $\ptw$ for which there exists only one critical point, and hence only one 
sonic point. The accretion is 
{\em mono-transonic} and the critical points are of a saddle type.
In  the region marked by 
{\bf I}, the critical points are called {\em inner} critical points since
these points are  quite close to the event horizon,  approximately  
in the range
$2<r_c^{\rm inner}{\le}10$. 
In the region marked by {\bf O}, the
critical points are called `outer critical points', because these points are
located considerably far away from the black hole. 
Depending on the value of
$\pthp$, an outer critical point may be as far as $10^4r_g$, 
or more.
The corresponding flow topologies are shown in the panel diagram
marked by {\bf I} 
and {\bf O}.
 
To describe the procedure for obtaining the panel plots, 
first consider  
the topology {\bf I}.
Using the value of $\pthp$ marked in the figure, we first solve Eq. (\ref{eq34})
 to obtain the 
corresponding critical point $r_c= 4.535$  marked 
in the figure by B  at the intersection of the accretion branch ABC and
the wind branch DBE. 
We then calculate the critical value of the advective 
velocity gradient at $r_c$ from Eq. (\ref{eq35})-(\ref{eq40}).
By integrating    (\ref{eq29}) and (\ref{eq31})
from the critical point B, using the fourth-order Runge-Kutta method, 
we then calculate the 
local advective velocity, the polytropic sound speed,
the Mach number, the fluid density, the disc height, the bulk temperature of the 
flow, and any other relevant dynamical and thermodynamic quantity
characterizing the flow.
In this way we  
obtain the accretion branch ABC by employing the above mentioned 
procedure.

Each solution represented in the panel plots in Fig. \ref{fig2} is two-fold
degenerate owing to the $\pm u$ degeneracy which
reflects the physical accretion/wind degeneracy. 
We have, however, removed the degeneracy by orienting the curves, and thus 
each line represents either the wind or accretion. 
 We have arbitrarily assigned  the $+$ sign solution in Eq. (\ref{eq35}) to the
  accretion
and the $-$ sign solution in Eq. (\ref{eq35}) to 
the `wind' branch DBE. This wind branch is just a 
mathematical counterpart of the accretion solution (velocity reversal 
symmetry of accretion), 
owing to the presence of the quadratic term 
of the dynamical velocity in the equation governing the 
energy momentum conservation. 

The term `wind solution' has 
a historical origin. 
The solar wind solution first introduced
by Parker  \cite{parker} 
has the same  topology profile as that of the 
wind solution obtained in classical Bondi accretion \cite{bondi}. Hence the 
name `wind solution' has been adopted in a more general sense.
The wind solution thus represents a hypothetical process,
in which, instead of starting from infinity 
and heading towards the black hole, the flow 
 generated near the black-hole event horizon would fly away from the
black hole towards infinity. The topology of such a process 
is represented by the wind solution DBE.

The above procedure for obtaining the flow topology 
is also applied to 
 draw the mono-transonic accretion/wind branch 
through the outer sonic point, i.e. the topology marked by 
{\bf O} with $r_c^{\bf outer}=318.63$. In fact, the same procedure may be 
used to draw real physical transonic accretion/wind solutions passing through any 
acceptable saddle-type critical point.
Note, however,  
that AB in the topology {\bf I} or {\bf O}, is {\it not} the complete
subsonic branch, because B is a critical point and not a sonic point.
Using the procedure described above 
we have to integrate the flow from B to the
sonic point 
where the Mach number equals one. The  sonic point is 
basically equivalent to the {\em acoustic horizon} at which 
the analogue Hawking radiation is emitted.
We  discuss this
issue in detail in 
sections \ref{sec:agahsg} and \ref{sec:atshbhad}.
Note
 that the regions 
{\bf I} or {\bf O} in the central plot 
do not include the `zero energy' accretion,
i.e. a flow parameterized by ${\cal E}=1$, although from the diagram it may 
so appear. Zero energy mono-transonic accretion does not 
admit a steady solution passing 
through the inner or the outer critical point.
Note also that the parameter space region {\bf I} extends
up to ${\cal E}=2$ along the horizontal axis, which is not shown 
in the figure for convenience.
\subsection{Mono-transonic solutions with two critical points \label{subsec:gcs_mts2cp}}
\noindent
The region marked by {\bf T} in the central plot, including the region marked by 
${\bf T}_1$ with 
${\ptw}_{\bf T_1}~{\subseteq}~{\ptw}_{\bf T}$,
represents $\ptw$ for which {\it two} critical points exist. 
We call these two critical points $r_c^1$ and $r_c^2$,
with $r_c^1<r_c^2$. The {\bf T} region is extended 
along the horizontal axis to the left 
up to $\lambda=0$, and along the vertical axis up to ${\cal E}=2$. 
A small patch of $\ptw|_{\bf T}$ not shown in the figure exists in the 
range $\left[1.0002{\le}{\cal E}{\le}1.0014,~5.18{\le}\lambda{\le}5.28\right]$.
Except this small patch, almost the whole region
in the $\ptw$ space beyond  the nearly vertical line 
passing through $\lambda=4$ generally does not admit any real solution for
$r_c$. 
The critical points obtained in the region 
$\ptw_{\bf T}$ can be further classified roughly into three regions for
${\cal E}>1$.

The first subset  $\ptw_{\bf T_1}$
is depicted in the  central plot as a dark triangular zone.
For this subset, both sets of critical points $r_c^1$ and $r_c^2$ lie within 
the radial distance of 10$r_g$, with   
$3.1{\le}r_c^1{\le}3.1014$ and 
$8.6{\le}r_c^2{\le}10$. We have found that $r_c^1$ is not associated with any 
steady solution passing through it, whereas there exists   a  complete 
mono-transonic accretion/wind solution passing through $r_c^2$ .
 However, an accretion solution belonging to this class 
is topologically not very different from 
 any mono-transonic solution obtained for $\ptwp{\in}{\ptw}_{\bf I}$
or $\ptwp{\in}{\ptw}_{\bf O}$.
 One such solution 
is shown in the panel figure marked by {\bf T}$_1$.
The critical point B for such solution is located at $r_c= 9.75$.
 By `complete' solution, we refer to a solution
that extends from $r=2$ to $r{\rightarrow}
{\infty}$.

The second subset of $\ptw_{\bf T}$ corresponds to the cases where $r_c^1$
ranges from 2 to 10, while $r_c^2$ is  
substantially larger.
Similarly, in this  region we do not find any 
steady solution passing through $r_c^1$ 
and we find a complete steady,  transonic
solution passing through $r_c^2$. 
Such a solution is shown in the side panel marked by {\bf T} with the critical 
point at about $r_c= 335$. 

The third subset of $\ptw_{\bf T}$, as mentioned earlier,
 is $\left[1.0002{\le}{\cal E}{\le}1.0014,~5.18{\le}\lambda{\le}5.28\right]$. 
Here $r_c^1$ ranges approximately from 60 to 115  and
$r_c^2$  approximately from 150 to 3500. 
Again, steady solutions passing through $r_c^1$ do not exist.
 However, the solutions
passing through $r_c^2$ are steady, but not complete, because those solutions 
form an outbound loop round the corresponding $r_c^1$ enclosed in the loop.
One such representative solution is shown in the side panel {\bf L}$_1$.
The critical point  
$r_c^2= 486.79$ is marked by B and the corresponding 
$r_c^1=64.08$ is marked by a crossed circle $\otimes$.

\begin{figure}
\includegraphics[scale=0.7,angle=270.0]{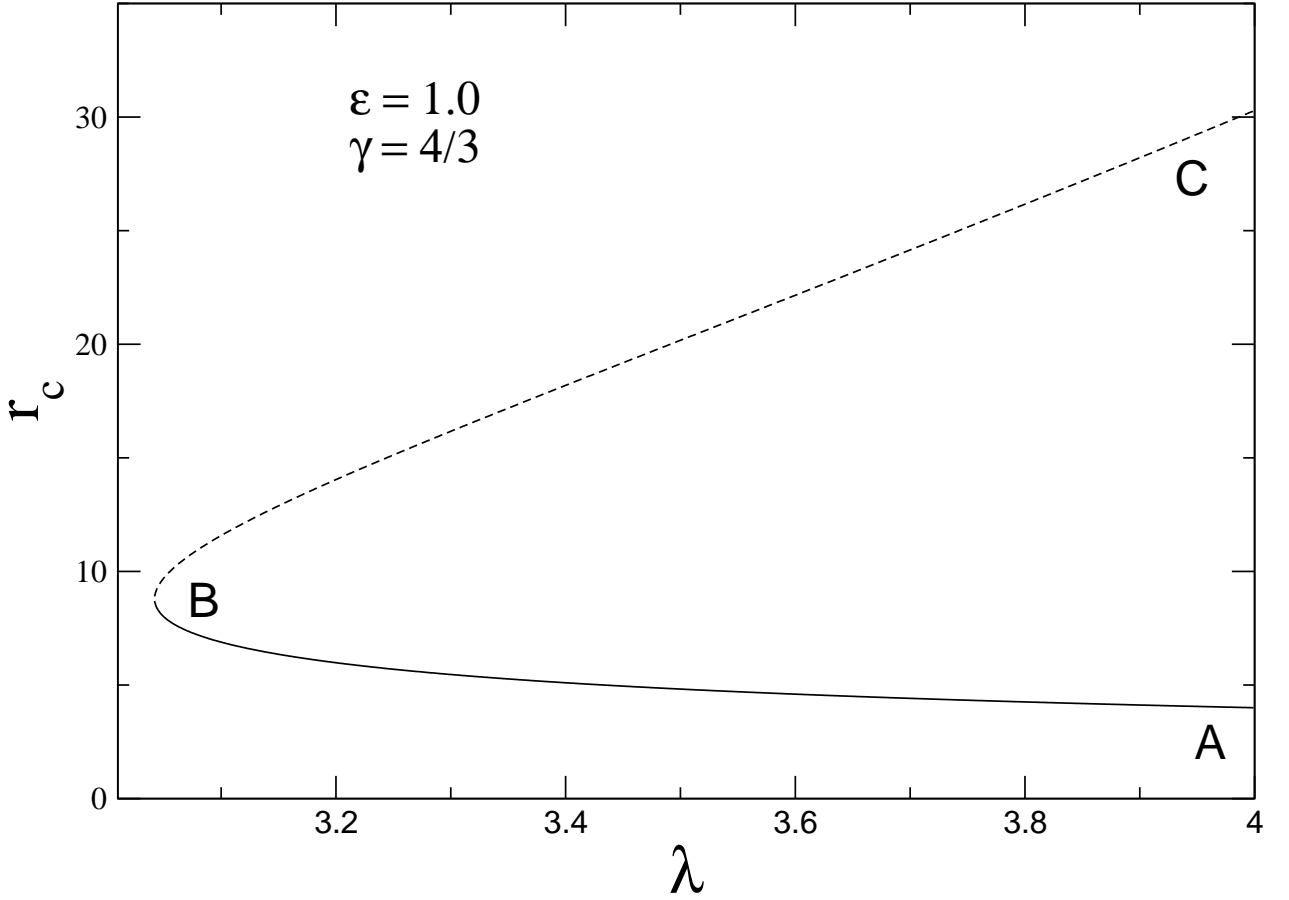}
\caption[h]{Zero energy (${\cal E}=1$) accretion 
flow with two critical points. The solid line AB represents the set of {\em inner} critical 
points. The flow passing through an inner critical
point has a closed loop topology as shown in the side panel {\bf L}$_2$
of Fig. \ref{fig2}. 
The dashed line represents the  set of {\em outer} critical points. 
There is no steady solution  passing through an outer critical point.}
\label{fig3}
\end{figure}

In the above discussion, we have concentrated on the solutions for which
${\cal E}>1$.
 In Fig. \ref{fig3} we plot  $r_c$ as 
a function of $\lambda$ for the zero energy accretion, i.e.
the accretion with ${\cal E}=1$.
The solid line AB represents $r_c^1$, whereas the dashed line 
represents $r_c^2$. 
We find no steady solutions passing through  $r_c^2$,
whereas there exists a solution
passing through $r_c^1$ which is smooth but incomplete as it forms an inward 
bound loop round $r_c^2$ enclosed in the loop. 
In the panel figure marked by {\bf L}$_2$, we plot one representative 
of such
topology. 
The critical point  
$r_c^1= 5.1$ is marked by B and the corresponding 
$r_c^2=18.19$ is marked by a crossed circle $\otimes$.
The curve
 ABCD represents the incomplete 
accretion (ABC)/wind (DBC) solution. Interestingly,
 this structure is topologically
similar to the inward bound loop representing the incomplete accretion/wind 
branch of a multi-transonic accretion
which we  discuss in the next  paragraph.

Note here that the accretion with two {\em critical}
points can practically be considered as an example of
mono-transonic accretion. Since one of the two critical
points (either $r_c^1$ or $r_c^2$) does not belong to any steady flow
through it, only the other critical point admits 
a steady flow to pass through it, and hence there will be only
one sonic point associated with it.
 
\subsection{Multi-transonic accretion and wind \label{subsec:gcs_mtaw}} 
\noindent
The wedge shaped region marked by {\bf A} in the central figure  corresponds to the 
multi-transonic accretion. 
For any $\ptwp{\in}{\ptw}_{\bf A}$,
the flow has three  critical points
 The corresponding flow topology is shown in 
the panel figure marked by {\bf A}. 
The line  ABC$_1$C$_2$D represents a complete
transonic solution passing through the saddle-type {\em outer} critical point B 
at $r_c=3310.88$.
The solid circle $\bullet$ marks the outer  sonic point at $r_s^{\rm outer} 
=2930.47$ where the flow Mach number is equal to one. The corresponding 
wind solution is shown by the line EBF.
 The crossed circle $\otimes$  
marks the location of the centre-type {\em middle} critical point at 
$r_c^{\bf middle}=
18.32$, through which a steady solution cannot be constructed.
The line EIGH represents the 
incomplete accretion (GIH)/wind (EIG) solution passing through the saddle type 
{\em inner} critical point I at $r_c=5.097 $.
The corresponding inner  sonic 
point $r_s^{\rm inner}$ (not shown in the figure) is located at $r_s=4.903$.
Clearly, 
this loop structure is very similar to the   zero energy accretion topology
 shown in the panel plot {\bf L}$_2$.

The set $\ptw_{\bf A}$ (or more  generally  $\pth_{\bf A}$) thus
produces doubly degenerate accretion/wind solutions.
 Such two 
fold degeneracy may be removed by the entropy considerations since 
the entropy rates ${\dot S}$($r_c^{\rm inner}$) and 
${\dot S}$($r_c^{\rm outer}$) are generally not equal.
 For any $\ptwp{\in}{\ptw}({\bf A})$, 
we find that the entropy rate  ${\dot S}$ evaluated for the 
complete accretion solution passing through the outer critical point
is {\it less} than that of the rate  evaluated for the incomplete accretion/wind solution
passing through the inner critical point.
 Since the quantity ${\dot S}$
is a measure of the specific entropy density of the flow, 
the solution passing through $r_c^{\rm outer}$ will naturally tend 
to make 
a transition to its higher entropy counterpart, 
i.e. the incomplete accretion solution 
passing through $r_c^{\rm inner}$. 
Hence, if 
there existed a mechanism for
the solution ABC$_1$C$_2$D 
to increase  
its entropy accretion rate by an amount 
\begin{equation}
{\Delta}{\dot S}=
{\dot S}(r_c^{\rm inner})-{\dot S}(r_c^{\rm outer}),
\label{eq48}
\end{equation}
there would be a transition to the
incomplete solution GIH.
 Such a transition would take place at 
a radial distance somewhere  between the radius of the inner  sonic point
 and the 
radius
of the accretion/wind turning point ($r= 94.9$ for this case)
marked by G.
 In this way one would obtain a combined accretion solution connecting 
$r{\rightarrow}{\infty}$ with $r=2$ which includes a part of the accretion 
solution passing through the inner critical, and hence the inner sonic point.
One finds that for some specific values of
$\pth{\subseteq}{\pth}_{\bf A}$, a standing Rankine-Hugoniot shock may accomplish this task. 
A supersonic accretion through the outer {\it sonic} point $r_s^{\rm outer}$ 
can generate
entropy through such a shock formation and can join the flow passing through 
$r_s^{\rm inner}$.
Although two shock locations, indicated by the vertical lines 
through C$_1$ and C$_2$ with a downward arrow, are found,
only one of the  
two shocks is stable. We discuss the details of such a shock formation 
in section \ref{sec:sfra}.

The wedge shaped region marked by {\bf W} in the central figure
 represents the $\ptw$ zone for which 
three critical points, the inner, the middle and the outer are also found. 
However, in contrast to $\ptw_{\bf A}$,
 the set $\ptw_{\bf W}$ yields  solutions 
 for which
  ${\dot S}(r_c^{\rm inner})$ is {\it less} than  
${\dot S}(r_c^{\rm outer})$. 
Besides,  
the topological flow profile of
 these solutions
is   different.
 One such solution topology is presented in the panel plot marked by {\bf W}.
Here the loop-like structure is formed through the {\it outer} critical point. 
Note a similarity with the topology {\bf L}$_1$.
This  topology  
is interpreted in the following way.
The flow HB passing through the inner critical point
B  at $r_c=4.25$
is the complete mono-transonic 
accretion flow, and 
ABC$_2$C$_1$D is its corresponding wind solution. The solution EFG passing through 
the outer critical point E at $r_c^{\rm outer}=301.53 $, with the corresponding 
outer  sonic point marked by a solid circle $\bullet$ located at 
$r_s^{\rm outer}= 343.312$
 represents the incomplete accretion (EF)/wind (FG) solution.
The initially 
subsonic wind solution passing through $r_c^{\rm inner}$ encounters 
the outer sonic point at 
$r_s^{\rm inner}= (4.31 $, and becomes supersonic.
 However, as 
${\dot S}_{\rm ABC_2C_1D}$ turns out to be less than  
${\dot S}_{\rm FEG}$, the solution branch ABC$_2$C$_1$D can make 
a shock transition to join 
its counter solution FEG and thereby
increase the entropy accretion rate by the amount
${\Delta}{\dot S}=
{\dot S}_{\rm  ABC_2C_1D}-{\dot S}_{\rm FEG}$. 
Here, two theoretical shock locations are 
obtained, out of which only one is stable. 
Hence the set $\pth_{\bf W}$ corresponds to  {\it mono-transonic} accretion
solutions  with
multi-transonic wind solutions with a shock.

Besides  $\gamma=4/3$, for which Fig. \ref{fig2}  has been drawn, we could perform 
a similar classification for any astrophysically relevant value of 
$\gamma$ as well. Some characteristic features of $\ptw$ would 
be changed as we vary $\gamma$. For example, if
${\cal E}_{\rm max}$ is the maximum value
of the energy and if $\lambda_{\rm max}$ and $\lambda_{\rm min}$ are the maximum and
the minimum values of the angular momentum, respectively,
for $\ptw_{\bf A}$ for a fixed value of
$\gamma$,
then
$\left[{\cal E}_{\rm max},\lambda_{\rm max},\lambda_{\rm min}\right]$ 
anti-correlates with $\gamma$.
Hence,  as the flow makes a transition
from its ultra-relativistic
to its purely non-relativistic limit,
the area representing $\ptw_{\bf A}$
decreases.
\begin{figure}[h]
\includegraphics[scale=1.5,angle=0.0]{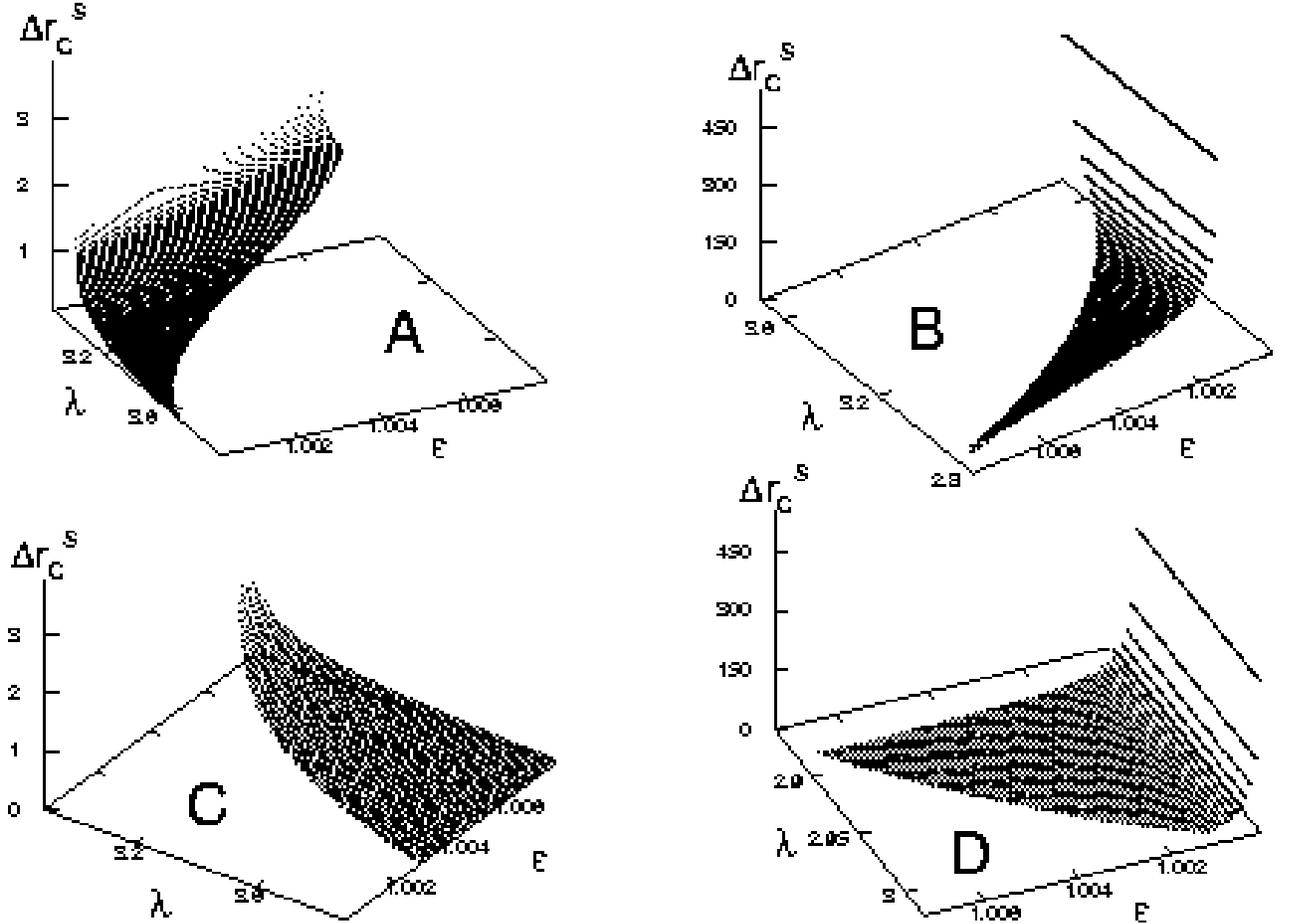}
\caption[]{The difference $\Delta{r_c^s}$ between of the sonic points radius
 $r_s$ and  the 
critical point radius $r_c$  as a function of the 
specific energy ${\cal E}$ and the specific angular momentum $\lambda$.
The plots represent
a mono-transonic flow passing through the inner 
critical point (plot A) and the outer critical point (plot B); a multi-transonic 
accretion passing through the inner critical point (plot C) and the outer critical 
point (plot D).}  
\label{fig4}
\end{figure}
\subsection{Dependence of $\Delta{r_c^s}$ on $\pth$ \label{subsec:gcs_deltar}}
\noindent
Here,  
we study the dependence of the radial difference between the
critical points and the sonic points on fundamental accretion parameters.
In Fig. \ref{fig4}, the difference ${\Delta}r_c^s$ defined by Eq. (\ref{eq46}) 
is plotted  as a function of the specific flow energy 
${\cal E}$  and the specific angular momentum $\lambda$
for a mono-transonic flow passing through the inner critical point (plot A) and
the outer critical point (plot B), and for a multi-transonic accretion flow
 passing through 
the inner critical point (plot C) and the outer critical point (plot D). 
Obviously,  
the radial difference between the sonic point and the critical point may be
quite significant, especially for the accretion/wind flow passing through the outer critical and sonic points.The quantity ${\Delta}r_c^s$  
anti-correlates with $\lambda$ for a general flow, whereas it correlates with 
${\cal E}$ for a flow passing through the inner critical point and 
anti-correlates with 
${\cal E}$  for a flow passing through the outer critical point. 

\subsection{Onset of chaotic behaviour in multi-transonic flow
\label{subsec:chaos}}
\noindent
In the central part of figure 2  the boundary between the {\bf O}
and the {\bf A} zone, and between the {\bf W} and the {\bf I}
zone, provides interesting information about the possibility of
the onset of chaotic behaviour of the multi-transonic accretion flow.
According to the theory of dynamical systems (see, e.g.
\cite{jordan}), 
if a small change of the control parameters leads
to profound effects on the outcome of the whole system, the system may then
be considered to exhibit chaotic behaviour. Along the boundary between
{\bf O} and {\bf A}, an extremely small change of ${\lambda}$
or ${\cal E}$ (or $\gamma$ if there were a transition 
 across the bounding
{\it surface} between $\left[{\cal E},\lambda,\gamma\right]_{\bf O}$
and $\left[{\cal E},\lambda,\gamma\right]_{\bf A}$) would lead to a sharp
transition from a mono-transonic  to a multi-transonic flow
(and vice versa). A similar situation arises for the
$\left[{\cal E},\lambda\right]_{\bf W}
{\longleftrightarrow}
\left[{\cal E},\lambda\right]_{\bf I}$ transition.
                                                                                                          
The above phenomenon is an example of the bifurcation of critical points due
to a slight change in the control parameters.
It is possible to study the onset of chaotic behaviour even for a
stationary system. Bray and Moore \cite{bray}, for example, studied 
the spin-glass phase in terms of a $T=0$ fixed point, and established
a criteria for the onset of chaos in such a system by proposing an effective
Lyapunov exponent. In a similar way we  propose 
an effective Lyapunov exponent for a transonic flow solution.
Let $u_1(r)$ and $u_2(r)$ be the values of 
the advective velocity at any 
radial distance $r$ for two transonic solutions 
characterized by
 very close values of the angular momentum $\lambda_1$ and 
$\lambda_2$ respectively , i.e. $ \left(\lambda_1-\lambda_2\right){\longrightarrow}0$,
with ${\cal E}$ and $\gamma$ kept fixed.
Let $u_1(r+R)$ and $u_2(r+R)$ be
 the corresponding velocities for the
above-mentioned transonic flow solutions
at the radial distance $r+R$.
We then define the
effective Lyapunov exponent $L_{eff}$, so that the 
following condition holds:
\begin{equation}
\frac
{u_1(r+R)-u_2(r+R)}
{u_1(r)-u_2(r)}
\sim
e^{L_{eff}R} .
\label{eqchao}
\end{equation}
The quantity $L_{eff}$ can also be calculated by varying ${\cal E}$ 
instead of 
$\lambda$. Next, one can show that the system will be sensitive to 
the initial boundary conditions if $L_{eff}>0$. The 
preliminary calculation \cite{naku} indeed indicates
 such values of $L_{eff}>0$
for a transition from mono- to multi-transonic flow, and vice versa.
Hence, transonic flow solutions  exhibit chaotic behaviour.
Investigation of chaos in black-hole accretion discs may shed a new light 
on the explanation of the variability mechanism of active 
galactic nuclei (for details about AGN variability, see, e.g.
\cite{milwit}). However, the details of the calculation are beyond 
the scope of this paper and will be presented elsewhere.

\section{Shock formation in relativistic accretion \label{sec:sfra}}
\subsection{A general overview \label{subsec:sfra_overview}}
\noindent
Perturbations of various kinds may produce discontinuities in
an astrophysical fluid flow.
By  {\em discontinuity} at a surface in a fluid flow we understand
any discontinuous change of
a dynamical or a thermodynamic quantity across  the 
surface. The corresponding surface is called a {\em surface of discontinuity}.
Certain boundary conditions must be satisfied across such surfaces and 
according to
these conditions, surfaces of discontinuities are classified into various categories.
The most important such discontinuities  
are  {\em shock waves} or {\em shocks}.
In an adiabatic flow of the Newtonian fluid, the shocks obey the following
conditions  \cite{landau}:
\begin{equation}
\left[\left[{\rho}u\right]\right]=0,~
\left[\left[p+{\rho}u^2\right]\right]=0,~
\left[\left[\frac{u^2}{2}+h\right]\right]=0,
\label{eq49}
\end{equation}
where $[[f]]$ denotes the discontinuity of $f$ across the surface of discontinuity, i.e.
\begin{equation}
\left[\left[f\right]\right]=f_2 -f_1,
\label{eq106}
\end{equation}
with $f_2$ and $f_1$ being the boundary values 
of the quantity $f$ on the two sides of
the surface.
 Such shock waves 
are quite often generated in
various kinds of supersonic astrophysical flows having
intrinsic angular momentum, resulting
in a flow which
becomes subsonic. This is because the repulsive centrifugal potential barrier
experienced by such flows is sufficiently strong to brake the infalling
motion and a stationary solution
could be introduced only through a shock. Rotating, transonic astrophysical fluid 
flows are thus believed to be `prone' to the shock formation phenomena.

One also
expects that a shock formation in black-hole accretion discs
might be a general phenomenon because shock waves
in rotating astrophysical flows potentially
provide an important and efficient mechanism
for conversion of a significant amount of the
gravitational energy  into
radiation by randomizing the directed infall motion of
the accreting fluid. Hence, the shocks  play an
important role in governing the overall dynamical and
radiative processes taking place in astrophysical fluids and
plasma accreting
onto black holes.
The study of steady, standing, stationary shock waves produced in black
hole accretion has acquired a very important status in recent
years. For  details
and for an exhaustive list of references 
see, e.g. \cite{apj1}.

Generally,
the issue of the formation of steady, standing shock waves in black-hole accretion discs is 
addressed in
two different ways. 
First, one can study the formation of Rankine-Hugoniot shock waves in a
polytropic flow. Radiative cooling in this type of shock is quite inefficient. No energy is
dissipated at the shock and the total specific energy of the accreting material is a shock-conserved
quantity. Entropy is generated at the shock and the post-shock flow possesses
 a higher entropy accretion rate
than its pre-shock counterpart. The flow changes its temperature permanently at the shock. Higher
post-shock temperature puffs up the post-shock flow and a quasi-spherical,
 quasi-toroidal centrifugal
pressure supported region is formed in the inner region of the accretion disc
\cite{apj1}.  

Another class of the shock studies concentrates on
the shock formation in isothermal black-hole accretion
discs. The characteristic features of such shocks are quite different from the
non-dissipative shocks discussed
above. In isothermal shocks, the
accretion flow  dissipates a part of its
energy and entropy at
the shock surface to keep the post-shock temperature equal to its pre-shock value.
This maintains the vertical
thickness of the flow exactly the
same just before and just after the shock is formed. Simultaneous jumps in
energy and entropy join the pre-shock supersonic flow to its post-shock
subsonic counterpart.
For detailed 
discussion 
and references
 see, e.g. \cite{apj2,tsuruta}.

In this work, the basic equations governing the flow 
are the energy and baryon number
conservation equations which contain no dissipative
terms and the flow is assumed to be inviscid.
Hence, the shock
which may be produced in this way can only be of Rankine-Hugoniot type
which conserves energy. The shock thickness must be very small
in this case, otherwise non-dissipative
flows may radiate energy through the upper and the lower boundaries because
of the presence of strong temperature gradient in between the inner and
outer boundaries of the shock thickness. 
In the  presence of a shock
the flow may experience the following profile.
A subsonic flow starting from infinity first becomes supersonic after crossing
the outer sonic point and somewhere in between the outer sonic point and the inner 
sonic point 
the shock transition takes place and forces the solution 
to jump onto the corresponding subsonic branch. The hot and dense post-shock
subsonic flow produced in this way becomes supersonic again after crossing
the inner sonic point and ultimately dives supersonically into the 
black hole.
A flow heading towards a neutron star can have the liberty of undergoing 
another shock transition
after it crosses the inner sonic point
\footnote{
Or, alternatively, a shocked flow heading towards a neutron star
need not have to encounter the inner sonic point at all.}, because the hard surface boundary
condition of a neutron star by no means prevents the flow 
from hitting the star surface subsonically.
\subsection{The relativistic Rankine-Hugoniot conditions \label{subsec:sfra_rRHc}}
\noindent
For the complete general relativistic accretion flow discussed in this work,
the energy momentum tensor ${T}^{{\mu}{\nu}}$, the four-velocity $v_\mu$,
and the speed of sound $c_s$ may have discontinuities at a 
hypersurface $\Sigma$ with its normal $\eta_\mu$.
 Using the energy momentum conservation and the
continuity equation, one has
\begin{equation}
\left[\left[{\rho}v^{\mu}\right]\right] {\eta}_{\mu}=0,
\left[\left[{T}^{\mu\nu}\right]\right]{\eta}_{\nu}=0.
\label{eq51}
\end{equation}
For a perfect fluid, one can thus formulate the relativistic
 Rankine-Hugoniot conditions as
\begin{equation}
\left[\left[{\rho}u\Gamma_{u}\right]\right]=0,
\label{eq52}
\end{equation}
\begin{equation}
\left[\left[{T}_{t\mu}{\eta}^{\mu}\right]\right]=
\left[\left[(p+\epsilon)v_t u\Gamma_{u} \right]\right]=0,
\label{eq53}
\end{equation}
\begin{equation}
\left[\left[{T}_{\mu\nu}{\eta}^{\mu}{\eta}^{\nu}\right]\right]=
\left[\left[(p+\epsilon)u^2\Gamma_{u}^2+p \right]\right]=0,
\label{eq54}
\end{equation}
where $\Gamma_u=1/\sqrt{1-u^2}$ is the Lorentz factor.
The first two conditions (\ref{eq52}) 
and (\ref{eq53}) 
 are trivially satisfied owing to the constancy of the
specific energy and mass accretion rate. 
The constancy of mass accretion yields
\begin{equation}
\left[\left[
K^{-\frac{1}{\gamma-1}}
\left(\frac{\gamma-1}{\gamma}\right)^{\frac{1}{\gamma-1}} 
\left(\frac{c_s^2}{\gamma-1-c_s^2}\right)^{\frac{1}{\gamma-1}}
uc_s
\sqrt{\frac{r^3-{\lambda^2}\left(r-2\right)}
{\gamma-\left(1+c_s^2\right)}}
\right]\right]=0.
\label{eq55}
\end{equation}
The third Rankine-Hugoniot condition 
(\ref{eq54})
may now be  written as
\begin{equation}
\left[\left[
K^{-\frac{1}{\gamma-1}}
\left(\frac{\gamma-1}{\gamma}\right)^{\frac{\gamma}{\gamma-1}}
\left(\frac{c_s^2}{\gamma-1-c_s^2}\right)^{\frac{\gamma}{\gamma-1}}
\left\{\frac{u^2\left(\gamma-c_s^2\right)+c_s^2}{c_s^2\left(1-u^2\right)}\right\}
\right]\right]=0.
\label{eq56}
\end{equation}
Simultaneous solution of Eqs. (\ref{eq55}) and (\ref{eq56}) yields the `shock invariant' 
quantity 
\begin{equation}
{\cal S}_h=
\frac{u^2\left(\gamma-c_s^2\right)+c_s^2}
{c_s{u}\left(1-u^2\right)\sqrt{\gamma-1-c_s^2}}
\label{eq57}
\end{equation}
which changes continuously across the shock surface.
We also define
 the {\em shock strength} ${\cal S}_i$ and the
{\em entropy enhancement} $\Theta$  as the ratio of the pre-shock
to post-shock Mach numbers (${\cal S}_i=M_{-}/M_{+}$),
 and as the ratio of the post-shock to pre-shock 
entropy accretion rates ($\Theta={\dot {S}}_{+}/{\dot {S}}_{-}$) of the 
flow, respectively.

\subsection{Shock locations in multi-transonic accretion and wind \label{subsec:sfra_slmaw}}
The shock location in 
multi-transonic accretion
is found in the following way.
Consider the multi-transonic 
flow topology {\bf A} in Fig. \ref{fig2}. 
Integrating along BC$_1$C$_2$D,  we calculate the shock invariant
${\cal S}_h$ in addition to
$u$, $c_s$ and $M$. We also calculate  ${\cal S}_h$ 
while integrating the sector HIG, starting from the inner {\it sonic}
point up to the point of inflexion G.
 We then determine
the radial distance $r_{sh}$, where the numerical values of ${\cal S}_h$,
obtained by integrating the two different sectors described above, are 
equal. Generally,
 for
any value of $\pthp$ allowing shock
formation,  one finds {\it two} shock locations
 marked by C$_1$ (the `outer'
shock -- between the outer and the middle
sonic points) and
C$_2$ (the `inner' shock -- between the
inner and the middle sonic points) in the figure.
 According to a standard
local stability analysis \cite{kafatos},
 for a multi-transonic accretion one can show that
only the shock formed  between
the middle
and the outer sonic point is stable.
Hence, in the multi-transonic accretion 
with the topology {\bf A},
the shock at  C$_1$ is stable and that
at  C$_2$ is unstable.
Hereafter, whenever we mention the shock 
location, we  refer
to the stable shock location only.

The topology {\bf W} in Fig. \ref{fig2} 
shows the shock formation in the multi-transonic {\it wind}.
Here the numerical values of ${\cal S}_h$ along the 
wind  solution passing through the inner sonic point (line BC$_2$C$_1$D)
are compared with the numerical values of ${\cal S}_h$
along the wind solution passing through the outer
sonic point, and the shock locations  $C_1$
and  $C_2$ for the wind are found accordingly.
\subsection{Geometry of the accretion with shock and generation of 
accretion-powered outflow \label{subsec:sfra_gsaapo}}
\begin{figure}
\includegraphics[scale=0.7,angle=270.0]{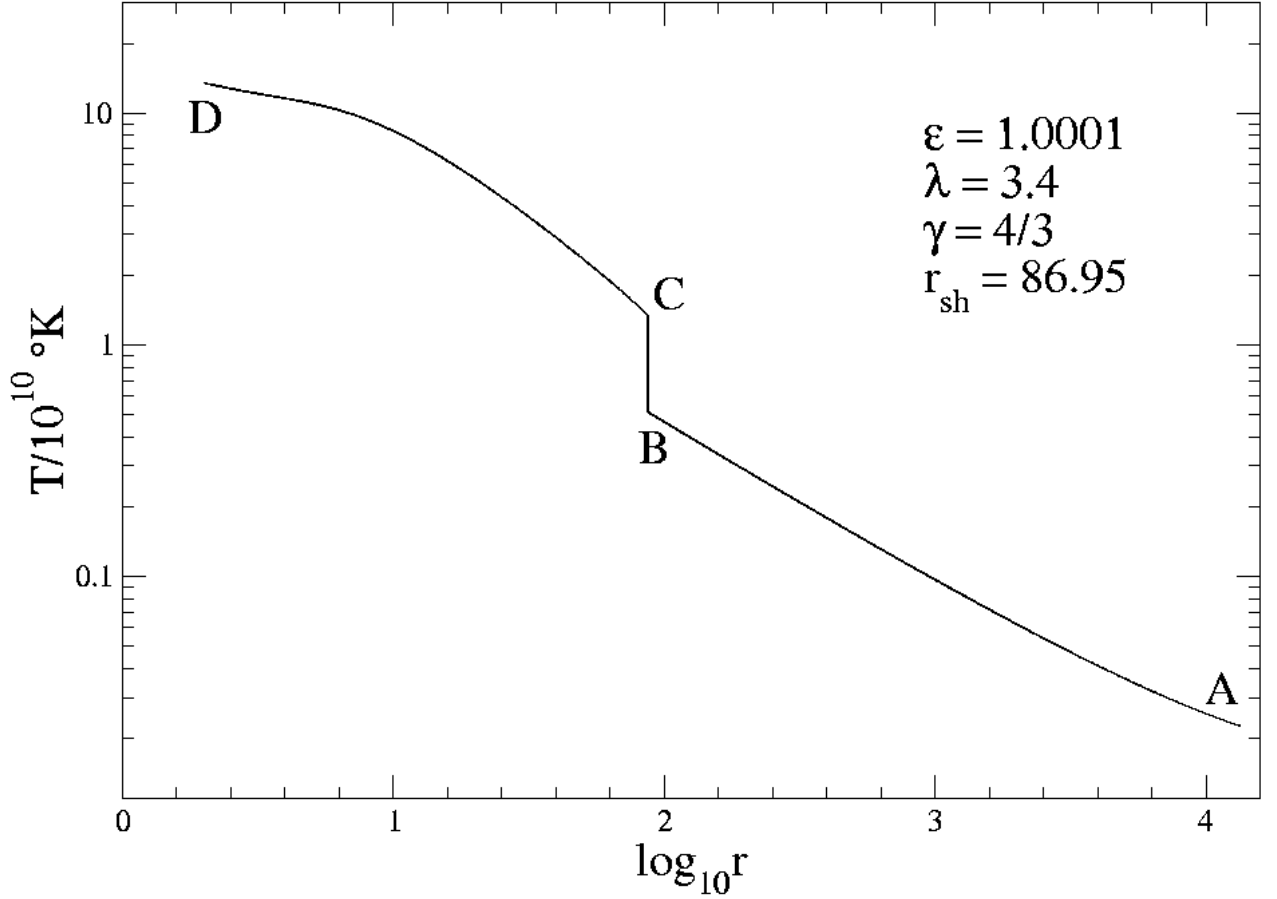}
\caption[]{Bulk temperature profile of a multi-transonic accretion 
flow for $\pthp$ shown with a shock at $r_{sh}$ . The vertical axis is the flow temperature  
in units of 10$^{10}$ K and the horizontal axis is the radial distance 
 in logarithmic scale. The sector AB represents the 
pre-shock flow,  the sector CD the post-shock flow, and the segment BC 
is the discontinuous increase of the flow temperature due to the 
shock formation.}
\label{fig5}
\end{figure}

\noindent
As a consequence of the shock formation in an accretion, 
the post-shock flow temperature will also increase abruptly.
In Fig. \ref{fig5}, we plot the temperature  of the combined accretion flow BC$_1$IH of the topology {\bf A} presented in Fig. \ref{fig2}.
 The segment AB corresponds to the pre-shock flow 
(BC$_1$ in the topology {\bf A} of Fig. \ref{fig2}), while the segment CD corresponds
to the post-shock flow. 
The vertical segment
BC is a discontinuous increase of the
flow temperature due to the shock formation. The length of BC, 
which measures the post- 
to pre-shock temperature ratio $T_+/T_-$ is, in general, 
a sensitive function of 
$\pth$.

\begin{figure}
\includegraphics[scale=0.65,angle=270.0]{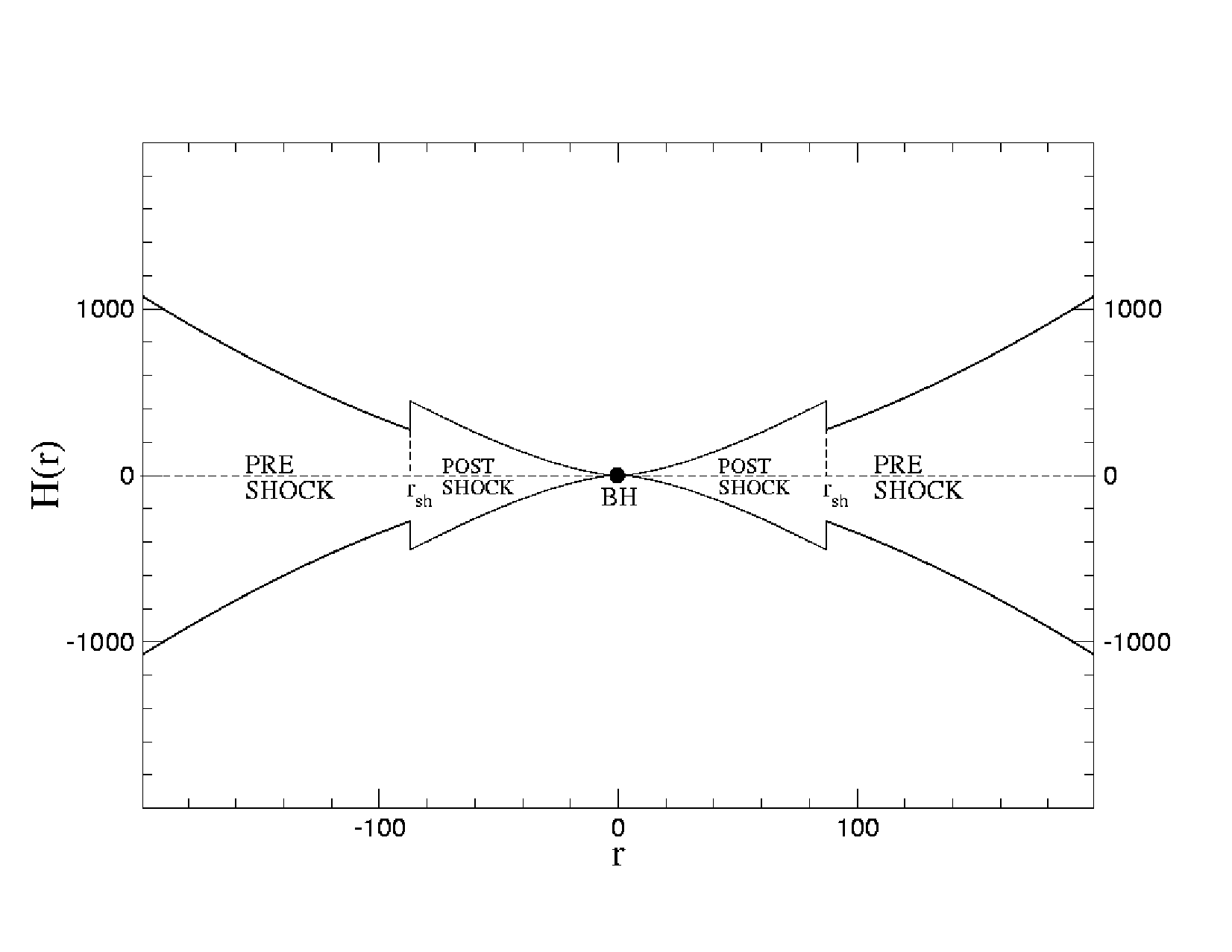}
\caption[]{Disc structure of  a multi-transonic accretion flow with a shock
at $r_{sh}$.
The disc 
height (vertical axis) is plotted as a function of the radial distance 
(horizontal axis). The point BH represents the black-hole event horizon.}
\label{fig6}
\end{figure}

In Fig. \ref{fig6}, we present the disc structure obtained by 
solving Eq. (\ref{eq28}) for the
combined accretion flow BC$_1$IH with the topology {\bf A} in Fig. \ref{fig2}.
The point BH represents the black-hole event horizon. The pre- and post-shock 
regions of the disc  are clearly distinguished in the figure 
and show 
that the post-shock disc puffs up significantly. 
It is clear from  Fig. \ref{fig5} that the bulk flow temperature
will be increased in the post-shock region.
Such an  increased disc temperature 
may lead to a disc evaporation resulting
in the formation of an optically thick halo.
Besides, 
 a strong temperature enhancement  may lead to the
formation of thermally driven outflows. 
  The generation of centrifugally 
driven and thermally driven outflows from black-hole accretion discs
 has been discussed
 in the
post-Newtonian framework \cite{dc99,santosh}.
The post-Newtonian approach
 may  be extended to general relativity using 
 the formalism  
 presented here.

Owing to the
 very high
 radial component of the
infall velocity of accreting material close to the black hole,
the viscous time scale is much larger than the infall time scale.
Hence, in the vicinity of the black hole,
a rotating inflow
entering the black hole will  have  an almost constant
specific angular momentum for any moderate viscous stress.
This angular momentum
yields a very strong centrifugal force  which
increases much faster than the gravitational force.
These two forces become comparable in size
 at
some specific radial distance.
At that point
 the matter starts
piling up and produces a boundary layer supported by the centrifugal pressure,
which may break the inflow to produce the shock. 
This actually happens  not quite at 
the point where the gravitational and centrifugal forces become equal but
slightly farther out
 owing to the thermal pressure.
Still closer to the black hole, gravity inevitably wins
and matter enters the horizon supersonically after passing
through a sonic point.
The formation of such a layer
may be attributed to the shock formation in accreting fluid.
The  post-shock flow becomes hotter and denser
and, for all practical purposes,
behaves as the stellar atmosphere as far as the formation of
outflows is concerned.
A part of the hot and dense shock-compressed in-flowing material
is then `squirted' as an outflow from the post-shock region.
Subsonic outflows originating
from the puffed up 
hotter post-shock accretion disc (as shown in the figure)
pass through the outflow sonic points and reach large distances
as in a wind solution.

The generation of such shock-driven outflows
is a reasonable assumption. A calculation describing the change
of linear momentum of the accreting material in the direction perpendicular to the
plane of the disc is beyond the scope of the disc model used in
this work because the explicit variation of dynamical variables along the Z axis 
(axis perpendicular to the equatorial plane of the 
disc)
cannot be treated
analytically.
The enormous post-shock thermal pressure
is capable of providing a substantial amount of `hard push' to the accreting
material against the gravitational attraction of the black hole. This `thermal
kick' plays an important role in re-distributing the linear momentum of the
inflow and generates a non-zero component along the Z direction. 
In other words,
the thermal pressure at the post-shock region,
being anisotropic in nature, may deflect a part of the inflow
perpendicular to the equatorial plane of the disc.
Recent work shows that \cite{monica} such shock-outflow model can 
be applied to successfully investigate the origin and dynamics of
the strong X-ray flares emaneting out from our galactic centre.

\subsection{Dependence of the shock location on $\pth$ \label{subsec:sfra_dsl}} 
\noindent
We find that the shock location  correlates with
$\lambda$.
This is obvious because the higher the flow
angular momentum, the greater the rotational energy content
of the flow. As a consequence, the strength of the centrifugal
barrier which is responsible to break the incoming flow by forming a shock
will be higher and
the location of such a barrier will be farther away from the
event horizon.
However, the shock location
anti-correlates with ${\cal E}$ and $\gamma$.
This  means that for the same ${\cal E}$ and $\lambda$, in the purely
non-relativistic flow the shock
will form closer to the black hole compared with
the ultra-relativistic flow. Besides, we find that the shock strength
${\cal S}_i$ anti-correlates with the shock location $r_{sh}$,
which indicates that
the
closer to the black hole the shock forms , the higher the strength ${\cal S}_i$
and the entropy enhancement ratio $\Theta$ are. The ultra-relativistic flows
are supposed to
produce the strongest shocks.
The reason behind this is also easy to understand. The closer to the black hole the shock
forms, the higher the available gravitational
potential energy must be released, and the radial
advective velocity required to have a more vigorous shock jump will be larger.
Besides we note  that as  the flow gradually approaches its purely
non-relativistic limit,
the shock may form for lower and lower angular momentum,
which indicates that for purely non-relativistic
accretion, the shock formation may take place even for a quasi-spherical flow.
However, it is important to mention that 
a shock formation will be allowed
not for every 
$\ptwp{\in}{\ptw}({\bf A})$
(or $\ptwp{\in}{\ptw}({\bf W})$) . 
Equation (\ref{eq57}) will be satisfied 
only 
for a specific subset of $\ptw({\bf A})$ (or $\ptw({\bf W})$),
for which a steady, standing shock solution
will be found.

\section{Acoustic geometry, acoustic horizons, and phonon quantization
\label{sec:agahsg}}
Before delving into the  calculation of $T_{AH}$ for various 
kinds of sonic horizon generated in the accretion disc, we would first like 
to discuss the relevant features of the acoustic geometry.

\subsection{Nonrelativistic acoustic geometry
 \label{subsec:agahsg_agfb} }
Let  $\psi$ denote 
 the velocity potential describing the fluid flow in Newtonian space-time, i.e.
 let
${\vec{u}}=-{\nabla}\psi$, where ${\vec{u}}$ is the velocity vector describing 
the dynamics of a Newtonian fluid. The specific enthalpy 
$h$ of a barotropic Newtonian fluid satsfies ${\nabla}h=(1/{\rho}){\nabla}p$, 
where $\rho$ and $p$ are the density and the pressure of the fluid. 
One then writes the Euler equation as
\begin{equation}
-{\partial}_t{\psi}+h+\frac{1}{2}\left({\nabla}{\psi}\right)^2+\Phi=0,
\label{eq58}
\end{equation}
where $\Phi$ represents the  potential  associated with any external driving force. 
Assuming small fluctuations around some steady background 
 $\rho^0, p^0$ and $\psi^0$, one 
can linearize the continuity and the Euler equations and 
obtain a wave equation
in the  form
(for derivation see appendix \ref{appendix})
\begin{equation}
{\partial}_{\mu}\left(f^{\mu{\nu}}{\partial}_{\nu}{\varphi}\right)=0 ,
\label{eq59}
\end{equation}
where ${\varphi}$ is  the first order term in  the expansion
$\psi=\psi^0+\epsilon \varphi+{\cal O}(\epsilon^2)$ and
the quantity
$f^{\mu{\nu}}$ in  this equation is a $4\times4$ symmetric matrix
defined by (\ref{eq009}).
Equation (\ref{eq59}) describes  the propagation 
of the linearized scalar potential $\varphi$. The function $\varphi$
represents the low amplitude fluctuations 
around the steady background 
 $(\rho^0,p^0,\psi^0)$ and 
thus describes the propagation  
of acoustic perturbances, .i.e. the propagation of 
sound waves.

The form of Eq. (\ref{eq59}) suggests that it may be 
regarded as  a d'Alembert  equation in curved space-time
geometry.
In any 
pseudo-Riemannian manifold the  d'Alembertia operator can be expressed as
\cite{mtw}
\begin{equation}
\Box =\frac{1}{\sqrt{-\left|g_{\mu{\nu}}\right|}}
{\partial}_{\mu}
{\sqrt{-\left|g_{\mu{\nu}}\right|}}
g^{\mu{\nu}}
{\partial}_\nu
\, ,
\label{eq61}
\end{equation}
where  $\left|g_{\mu{\nu}}\right|$ is the 
determinant  and 
 $g^{\mu{\nu}}$ is the  inverse of the metric
 $g_{\mu{\nu}}$.
Next, if one identifies
\begin{equation}
f^{\mu{\nu}}=\sqrt{-\left|g_{\mu{\nu}}\right|}g^{{\mu}\nu} ,
\label{eq62}
\end{equation}
 one can recast the acoustic wave equation 
in the form \cite{visser}
\begin{equation}
\frac{1}{\sqrt{-\left|G_{\mu{\nu}}\right|}}
{\partial}_\mu
\left(
{\sqrt{-\left|G_{\mu{\nu}}\right|}}
G^{\mu{\nu}}
\right){\partial_\nu}\varphi=0 ,
\label{eq63}
\end{equation}
where $G_{\mu{\nu}}$ is the acoustic metric tensor for the 
Newtonian fluid. The explicit form of ${G}_{\mu{\nu}}$
is obtained as
\begin{equation}
G_{\mu\nu}
\equiv {\rho^0}
\left[ \matrix{-(c_s^2-u^2)&\vdots&-{{\vec u}}\cr
               \cdots\cdots\cdots\cdots&\cdot&\cdots\cdots\cr
               -{\vec u}&\vdots& {I}\cr } \right]
\label{eq64}
\end{equation}
Thus, the propagation of acoustic perturbation, or the sound 
wave, embedded in a barotropic, irrotational, non-dissipative Newtonian fluid 
flow may be described by a scalar d'Alembert equation in a {\it curved} acoustic 
geometry. The corresponding  acoustic metric tensor is a matrix that 
depends on dynamical and thermodynamic variables parameterizing the fluid flow.

The acoustic metric  (\ref{eq64}) in many aspects resembles 
a black-hole type geometry
in general relativity. For example, the notions such as 
`ergo region' and 
  `horizon' may be introduced in full analogy with those of 
  general relativistic
 black holes.
For a stationary flow, the time translation Killing vector 
$\xi\equiv\partial/\partial t$ 
leads to the concept of 
{\em acoustic ergo sphere} as a surface at which
$G_{\mu\nu}\xi^\mu\xi^\nu$ changes its sign.
 The acoustic ergo sphere is  the envelop of the {\em acoustic ergo region} 
where $\xi^{\mu}$ is space-like with respect to the acoustic metric. 
Through the equation
$G_{\mu\nu}\xi^\mu\xi^\nu =g_{tt}=u^2-c_s^2$,
it is obvious that 
inside the ergo region  the fluid is 
supersonic. 
The `acoustic horizon' 
can be defined  as the boundary of a region from which 
acoustic null geodesics 
or phonons, cannot escape. Alternatively, the acoustic horizon is defined as a 
time like hypersurface defined by the equation
\begin{equation}
c_s^2-u_{\perp}^2=0 ,
\label{eq65}
\end{equation}
where $u_{\perp}$ is the component of the fluid velocity perpendicular to the acoustic horizon. Hence, any steady supersonic flow described in a stationary geometry by a time independent velocity vector field forms an ergo-region, inside 
which the acoustic horizon is generated at those points 
where the normal component of the fluid 
velocity is equal to the speed of sound.

\subsection{Acoustic geometry in a curved space-time \label{subsec:agahsg_agcb}}
The above formalism may be extended to relativistic fluids in curved space-time 
background \cite{bilic}. The propagation of acoustic disturbance in a 
perfect relativistic inviscid irrotational fluid
is  also described  by the wave equation of the form
(\ref{eq63})
in which
 the acoustic metric tensor and its inverse are defined as
\begin{equation}
{G}_{\mu\nu} =\frac{\rho}{h c_s}
\left[g_{\mu\nu}+(1-c_s^2)v_{\mu}v_{\nu}\right];
\;\;\;\;
G^{\mu\nu} =\frac{h c_s}{\rho}
\left[g^{\mu\nu}+(1-\frac{1}{c_s^2})v^{\mu}v^{\mu}
\right] ,
\label{eq72}
\end{equation}
where $\rho$ and $h$ are, respectively,
the rest-mass density and the specific enthalpy of the relativistic fluid,
$v^{\mu}$ is the four-velocity,
and $g_{\mu\nu}$ the background space-time metric. 
The ergo region is again defined as the region where the stationary 
Killing vector $\xi$ becomes spacelike and the acoustic horizon 
 as a timelike hypersurface the wave velocity of which
  equals the speed of sound at every point. The defining equation  
for the acoustic horizon is again of the form  (\ref{eq65}) 
in which the three-velocity component  perpendicular to the horizon
is given by
\begin{equation}
u_{\perp}=\frac
{
\left(\eta^\mu v_\mu\right)^2}
{
\left(\eta^\mu{v_\mu}\right)^2 +\eta^\mu\eta_\mu},
\label{eq73}
\end{equation}
where $\eta^\mu$ is the unit normal to the horizon.

It may be shown that, for an axisymmetric flow, the acoustic metric
 discriminant defined as 
\begin{equation}
{\cal D}=G_{t\phi}^2-G_{tt}G_{\phi\phi}
\label{eq81}
\end{equation}
vanishes at the acoustic horizon.
A supersonic flow
is characterized by the condition ${\cal D}>0$, whereas for a subsonic flow,  
 ${\cal D}<0$  \cite{abraham}.   According to the classification of Bercelo 
et al.\ \cite{berceloetal},  a transition from a subsonic (${\cal D}<0$)
 to a supersonic (${\cal D}>0$) flow
is an acoustic {\em black hole},
whereas  a transition from a
supersonic to a subsonic flow is an acoustic 
{\em white hole}.
 Hence, for the relativistic disc geometry 
presented in our paper, one can  show that for multi-transonic
shocked accretion and wind, an acoustic white hole produced at the shock
location $r_{sh}$ is flanked by two acoustic black holes produced
at the inner  and the outer  sonic points. 
The situation is precisely the same as in
the accretion geometry with a constant thin disc height 
considered in
\cite{abraham}  where a detailed description of the 
emergence of such black or  white holes and of the corresponding causal structures
is given.

\subsection{Quantization of phonons and the Hawking effect
\label{subsec:quant}}
The purpose of this section is to demonstrate how the
quantization of phonons in the  presence of
the acoustic horizon yields
acoustic Hawking radiation.
The acoustic perturbations considered here are classical
sound waves or {\em phonons} that
satisfy the massles wave 
equation 
in curved background, i.e.  
the general relativistic analogue of
(\ref{eq63}),
with the metric $G_{\mu\nu}$
given by (\ref{eq72}).
Irrespective of the underlying microscopic structure
acoustic perturbations are quantized.
A precise quantization scheme for an analogue gravity
system may be rather involved \cite{stuz}.
However, at the scales larger than
the atomic scales below which a perfect fluid description breaks down,
the atomic substructure may be neglected and
the field may be considered elementary. Hence,
the quantization proceeds in the same way as in the case of a
scalar field in curved space \cite{bir}
with a suitable UV cutoff for the scales below a typical
atomic size of a few \AA. 

For our purpose, the most convenient
quantization prescription is the Euclidean path integral
formulation.
Consider a 2+1-dimensional disc geometry.
The equation of motion (\ref{eq63}) with (\ref{eq72}) 
follows from the  variational principle applied to
the action functional 
\begin{equation}
S[\varphi]=\int dtdrd\phi\, 
\sqrt{-G}\,
G^{\mu\nu}
\partial_{\mu}\varphi
\partial_{\nu}\varphi\, .
\label{eq228}
\end{equation}
We define the functional integral 
\begin{equation}
Z= \int {\cal D}\varphi e^{-S_{\rm E}[\varphi]} ,
\label{eq229}
\end{equation}
where $S_{\rm E}$ is the Euclidean  action 
obtained from (\ref{eq228}) by setting
$t=i\tau$
and continuing the Euclidean time $\tau$ from imaginary to real values.
For a field theory at zero temperature, the integral
over $\tau$ extends up to infinity.
Here,
 owing to the presence of the acoustic horizon,
 the integral over $\tau$ 
 will be cut at the inverse Hawking temperature $2\pi/\kappa$
 where $\kappa$ denotes the analogue surface gravity.
 To illustrate how this happens, consider, for simplicity, a non-rotating 
 fluid ($v_\phi=0$) in the Schwarzschild  space-time.
 It may be easily shown that the acoustic metric takes the form
 \begin{equation}
ds^2=g_{tt}\frac{c_s^2-u^2}{1-u^2}dt^2 -2u\frac{1- c_s^2}{1-u^2}
drdt-\frac{1}{g_{tt}}\frac{2-c_s^2u^2}{1-u^2}dr^2 +r^2d\phi^2\, ,
\label{eq232}
\end{equation}
where $g_{tt}=-(1-2/r)$, $u=|v_r|/\sqrt{-g_{tt}}$,
and we have omitted the irrelevant conformal factor $\rho/(hc_s)$.
Using the coordinate transformation
\begin{equation}
dt\rightarrow dt+\frac{u}{g_{tt}}\frac{1- c_s^2}{c_s^2-u^2}dr
\label{eq233}
\end{equation}
we remove the off-diagonal part from (\ref{eq232}) and obtain
\begin{equation}
ds^2=g_{tt}\frac{c_s^2-u^2}{1-u^2}dt^2 -
\frac{1}{g_{tt}}\left[\frac{2-c_s^2u^2}{1-u^2}+
\frac{u^2(1- c_s^2)^2}{(c_s^2-u^2)(1-u^2)}\right]dr^2 
 +r^2d\phi^2.
\label{eq234}
\end{equation}
Next, we evaluate the metric near the acoustic horizon at
$r=r_{\rm s}$ using the expansion in $r-r_{\rm s}$ at first order
 \begin{equation}
c_s^2-u^2\approx 2 c_s \left. \frac{\partial}{\partial r}(c_s-u)
\right|_{r_{\rm s}}(r-r_{\rm s})
\label{eq235}
\end{equation}
and making the substitution
\begin{equation}
r-r_{\rm s}=\frac{-g_{tt}}{2c_s (1-c_s^2)}
\left. \frac{\partial}{\partial r}(c_s-u)\right|_{r_{\rm s}} R^2,
\label{eq236}
\end{equation}
where $R$ denotes a new radial variable.
Neglecting the first term in the square brackets in (\ref{eq234})
and setting $t=i\tau$, we obtain the Euclidean metric in the form
\begin{equation}
ds_{\rm E}^2=\kappa^2 R^2 d\tau^2 +dR^2 +r_{\rm s}^2 d\phi^2\, ,
\label{eq437}
\end{equation}
where 
\begin{equation}
\kappa=\frac{-g_{tt}}{1-c_s^2}
\left|\frac{\partial}{\partial r}(u-c_s)\right|_{r_{\rm s}}\, .
\label{eq238}
\end{equation}
Hence, the metric near $r=r_{\rm s}$ is the product of the metric on S$^1$
and the Euclidean Rindler space-time 
\begin{equation}
ds_{\rm E}^2=dR^2 + R^2 d(\kappa \tau)^2 .
\label{eq239}
\end{equation}
With the periodic identification 
$\tau\equiv \tau+2\pi/\kappa$, the metric (\ref{eq239})
describes $\mathbb R^2$ in plane polar coordinates. 

Furthermore, making the substitutions
$R=e^{\kappa x}/\kappa$ and $\phi=y/r_{\rm s}+\pi$,
the Euclidean action takes the form of the 
2+1-dimensional free scalar field action
at non-zero temperature
\begin{equation}
S_{\rm E}[\varphi]=\int_0^{2\pi/\kappa} d\tau 
\int_{-\infty}^{\infty}dx
\int_{-\infty}^{\infty} dy \frac{1}{2} (\partial_{\mu} \varphi)^2,
 \label{eq240}
\end{equation} 
where we have set
the upper and lower bounds of the integral over $dy$
to $+\infty$ and $-\infty$, respectively,
assuming  that $r_{\rm s}$ is sufficiently large.
Hence, the functional integral $Z$ in (\ref{eq229}) 
is evaluated over the fields $\varphi(x,y,\tau)$ that are periodic in
$\tau$ with period  $2\pi/\kappa$.
In this way, the functional $Z$ is just the
 partition function for a grandcanonical ensemble of free  bosons
  at the 
Hawking temperature
$T_{\rm H}=\kappa/(2\pi\kappa_B)$.
However, the radiation spectrum will not be exactly thermal
since we have to cut off the scales below the atomic scale
\cite{95unruh}. The choice of the cutoff and the deviation of 
the acoustic radiation spectrum from the thermal spectrum is
cloasely related to the so-called {\em transplanckian problem}
of Hawking radiation \cite{jacc}.

In the  Newtonian approximation, 
equation
(\ref{eq238}) reduces to the usual
non-relativistic expression for the acoustic surface gravity 
\cite{visser}
\begin{equation}
\kappa=
\left|\frac{\partial}{\partial r}(u-c_s)\right|_{r_{\rm s}}\, .
\label{eq241}
\end{equation}

\section{Analogue temperature of sonic horizons in a black-hole accretion disc 
\label{sec:atshbhad} }
In general, the ergo-sphere and the acoustic horizon do not coincide. However,
for some specific stationary geometry they do. This is the case, e.g.
 in the following two 
examples:

\begin{enumerate}
\item
Stationary spherically symmetric configuration where
fluid is radially falling into a pointlike drain at the origin. Since
$u=u_{\perp}$ {\em everywhere}, there will  be no
 distinction between the ergo-sphere and the acoustic horizon. 
An astrophysical example of such a situation is the
stationary  
spherically symmetric Bondi-type accretion \cite{bondi}
 onto a 
Schwarzschild  black hole. 

\item
Two-dimensional axisymmetric configuration, where the fluid is 
radially  
moving towards a drain placed at the origin.  Since only 
the radial component of the  velocity is non-zero,
$u=u_{\perp}$ everywhere. Hence, for this system, the acoustic horizon will 
coincide with the ergo region.
An astrophysical example is an axially symmetric 
accretion with zero angular momentum onto a Schwarzschild black 
hole or onto a non-rotating neutron star.
\end{enumerate}
 
Here we consider the 
axisymmetric accretion in which the ergo-sphere and the acoustic horizon do not coincide. 
 For a transonic  
accretion disc, the collection of sonic points at a fixed radial 
distance forms a surface, the generators of which are the trajectories of phonons, 
because at those points
$u_{\perp}=c_s$. This surface is an acoustic event horizon.

The analogue surface gravity at the acoustic horizon can be calculated as 
a function of fundamental accretion parameters, i.e. as a function of $\pth$.
In complete analogy to the classical black-hole horizon,
the acoustic horizon emits the  
analogue Hawking radiation consisting of thermal phonons.
The temperature of the analogue  Hawking radiation  $T_{AH}$ is related to
the analogue surface gravity as usual \cite{visser}. 
An axisymmetric transonic accretion disc round 
an astrophysical black hole is thus  a natural example of
analogue gravity model where two types of horizon (the gravitational 
and the acoustic), exist simultaneously.

For a stationary configuration, the surface gravity can be computed in terms 
of the Killing vector 
\begin{equation}
\chi^{\mu}=\xi^{\mu}+\Omega\phi^{\mu} 
\label{eq175}
\end{equation}
that is null at the acoustic horizon.
Following the standard procedure \cite{bilic,wald} one finds that 
the expression 
\begin{equation}
\kappa\chi^\mu=
\frac{1}{2}
G^{\mu\nu}\eta_\nu\frac{\partial}{\partial{\eta}}
(G_{\alpha\beta}\chi^{\alpha}\chi{\beta})
\label{eq75}
\end{equation}
 holds at the acoustic horizon.
 Here
    the constant $\kappa$ is  the surface gravity, 
 the vector $\eta_{\mu}$ is the unit normal to the
acoustic horizon and
$\partial/{\partial}\eta$ denotes the normal derivative 
\begin{equation}
\frac{\partial}{\partial{\eta}}\equiv\eta^\mu\partial_\mu=
\frac{1}{\sqrt{g_{rr}}}\frac{\partial}{\partial r}\, .
\label{eq176}
\end{equation} 
  From equation (\ref{eq75}) we
deduce the magnitude of the surface gravity  as
 \cite{bilic}
\begin{equation}
\kappa=
\left|\frac{\sqrt{-{\chi}^{\nu}{\chi}_{\nu}}}{1-c_s^2}
\frac{\partial}{\partial{\eta}}
\left(u-c_s\right)\right|_{\rm r=r_s}  ,
\label{eq76}
\end{equation}
where $r_s$ denotes the  location of the acoustic horizon.

For the transonic disc geometry described in section \ref{sec:tbhadgr}, we calculate the 
norm of $\chi^\mu$
at the acoustic horizon as
\begin{equation}
\left.\sqrt{-\chi^\mu\chi_\nu}\right|_{\rm r=r_s}
=r_s^{-2}\sqrt{\left(r_s-2\right)\left(r_s^3-\lambda^2r_s+2\lambda^2\right)}.
\label{eq77}
\end{equation}
The surface gravity is thus obtained as
\begin{equation}
\kappa=
 \left(1-\frac{2}{r_s}\right)
\sqrt{1-\frac{\lambda^2}{r_s^2}+\frac{2\lambda^2}{r_s^3}}\,
\left|
\frac{1}{1-c_s^2}
\frac{d}{dr}
\left(u-c_s\right)
\right|_{\rm r=r_s} ,
\label{eq78}
\end{equation}
with the corresponding analogue Hawking temperature given by
\begin{equation}
 T_{AH}=
\frac{\kappa}{2\pi \kappa_B} \, ,
\label{eq79}
\end{equation}
where we have used the units in which $\hbar=c=1$.

It is now easy to 
calculate $T_{AH}$ for each kind of 
sonic points 
(classified according to 
section \ref{sec:gcs}) as a function of $\pth$.
 We now define the quantity $\tau$ as the 
ratio of the analogue to the black-hole Hawking temperature
\begin{equation}
\tau=\frac{T_{AH}}{T_H}\, .
\label{eq80}
\end{equation}
This ratio turns out to be independent of the 
black-hole mass. Hence, we are able to compare the properties 
of the analogue and black-hole horizons for an accreting black hole with any mass, 
ranging  from the primordial black holes in the early Universe
 to the
supermassive black holes at galactic centres.

\begin{figure}[h]
\includegraphics[scale=1.5,angle=0.0]{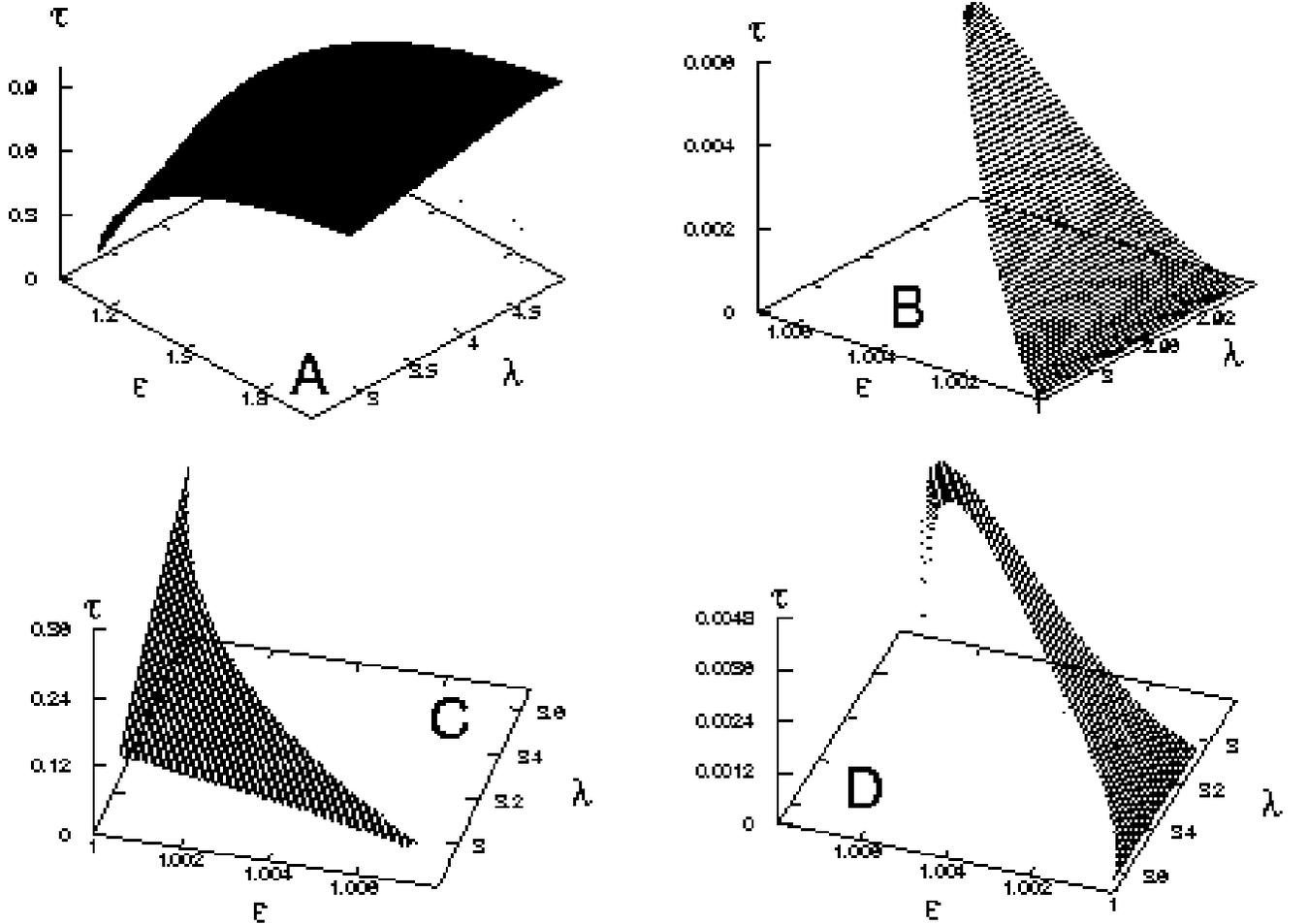}
\caption[]{The ratio of the analogue to Hawking temperature  
 $\tau$ (vertical axis) as a function of the specific energy of the 
flow ${\cal E}$ and the specific angular momentum $\lambda$. The  
representative cases  shown are: mono-transonic flow passing through the
single inner type sonic point (A), mono-transonic flow passing through the
single outer type sonic point (B), multi-transonic accretion passing 
through the inner sonic point (C), and multi-transonic accretion passing 
through the outer sonic point (D).}
\label{fig7}
\end{figure}

In Fig. \ref{fig7}, we plot the value of $\tau$  as a function of the 
specific energy ${\cal E}$ 
and the specific angular momentum $\lambda$ 
of the flow.
Four representative cases are shown:
\begin{description}
\item[A    ]  Mono-transonic flow passing through the 
single inner type sonic point. The range of $\ptwp$  
here corresponds to the central
region of Fig. \ref{fig2} marked by {\bf I}, except that ${\cal E}$ 
is extended up to 2.
\item[B    ]  Mono-transonic flow passing through the
single outer type sonic point. The range of$\ptwp$ used
to obtain the result for this region corresponds to the central
region of Fig.\ref{fig2} marked by {\bf O}.
\item[C    ] Multi-transonic accretion passing through the 
inner sonic point.  The range of $\ptwp$  corresponds to the central
region of Fig. \ref{fig2} marked by {\bf A}
\item[D    ] Multi-transonic accretion passing through the
outer sonic point. The range of $\ptwp$  corresponds to the central
region of Fig. \ref{fig2} marked by {\bf A}
\end{description}
Similar figures 
can be drawn for other types of transonic flow  for any
value of $\pthp$ allowing a real physical transonic solution.

It is obvious from  Fig. \ref{fig7} that for $\gamma=4/3$, the temperature $T_{AH}$ 
asymptotically approaches $T_H$ and is never higher than $T_H$.
However, we observe 
that $\tau$ increases non-linearly with $\gamma$, and at 
very high values of $\gamma$, e.g. $\gamma=1.63$, 
one does obtain $\ptw$ (mainly for high values of ${\cal E}$)
for which the analogue Hawking temperature {\em exceeds} the 
black-hole Hawking temperature. 

\begin{figure}
\includegraphics[scale=0.7,angle=270.0]{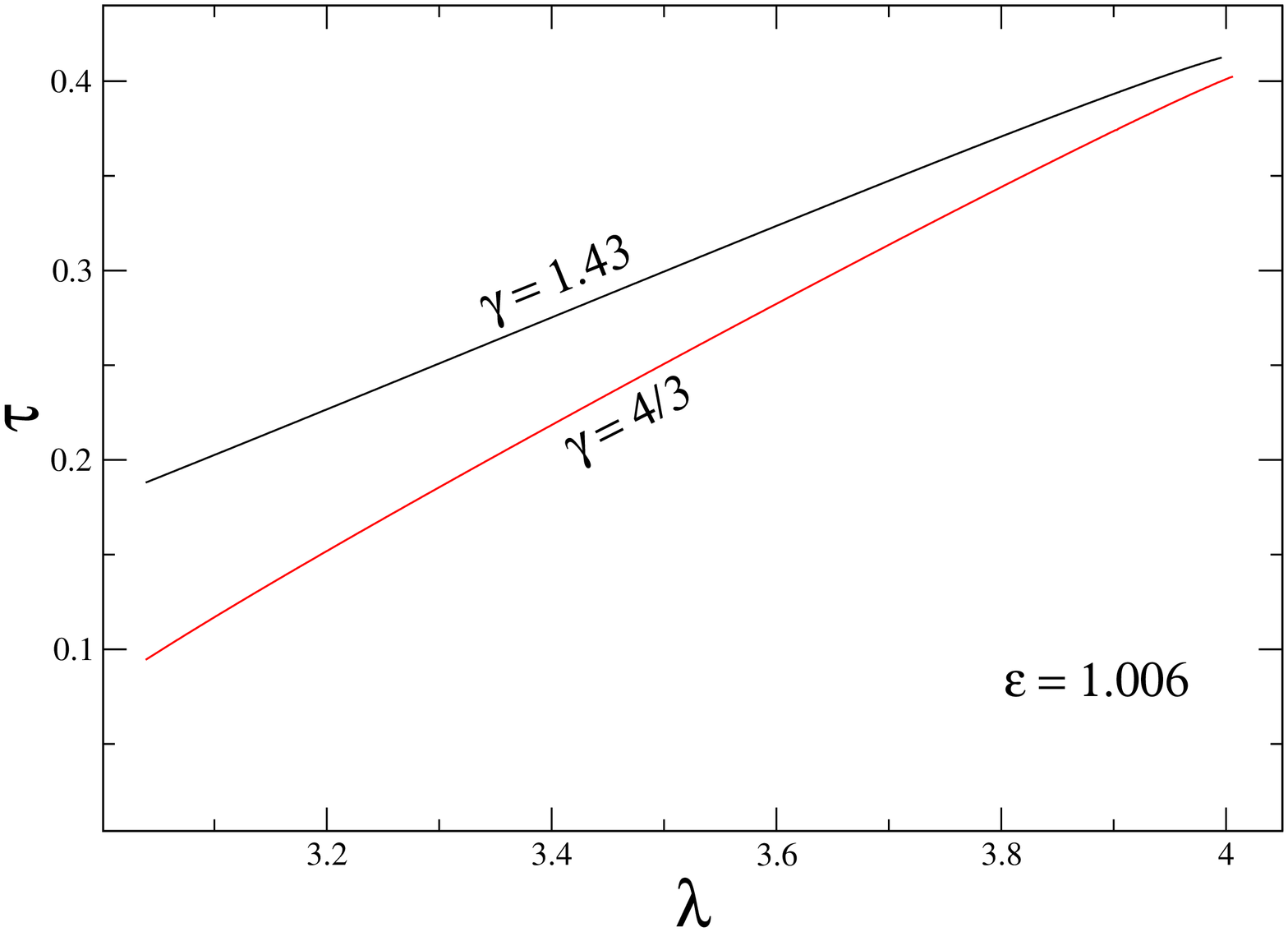}
\caption[]{
The ratio of the analogue to Hawking temperature $\tau$ as a 
function of specific angular momentum $\lambda$ for for a fixed specific energy ${\cal E}$ and two values 
of polytropic index $\gamma$.}
\label{fig8}
\end{figure}

In Fig. \ref{fig8}, we plot the variation of 
$\tau$ with $\lambda$ for a fixed ${\cal E}$ for two 
values of $\gamma$. Obviously, 
the ratio of the analogue to the black-hole Hawking 
temperature keeps increasing with $\gamma$.

\section{Discussion \label{sec:discussion}}
\noindent
In this work, we have established that the general relativistic axisymmetric 
accretion disc round an astrophysical black hole can be considered as an 
example of classical analogue gravity realized in nature.
To accomplish this task, we have first formulated and solved the equations describing 
the general relativistic axisymmetric accretion flow round a compact object.
 We then show 
that such accretion is transonic, and the collection of sonic points
forms an acoustic horizon.
The acoustic horizon is characterized by  analogue surface 
gravity and the corresponding analogue Hawking radiation . We have calculated the corresponding 
analogue temperature as a function of fundamental accretion parameters.
 We have 
shown that for some region of parameter space spanned by the specific energy ${\cal E}$, the
specific angular momentum $\lambda$ and the adiabatic index of the flow $\gamma$, 
the analogue Hawking temperature may {\it exceed} the  Hawking temperature of the 
accreting black hole.
Generally, this happens for high values of ${\cal E}$ and $\gamma$.

Note that the formalism presented in this work is not restricted to
analogue effects in the black hole accretion only. Hydrodynamic
accretion onto any astrophysical object, e.g. a weakly magnetized
neutron star, if it happens to be transonic, will produce
an acoustic horizon, and thus the analogue effects in such systems
can be studied using the methodology developed here.
Thus, we show in general, that a hydrodynamic transonic
accretion in astrophysics may be regarded as a natural example of
analogue gravity model.

It is important to
note that the accreting astrophysical black holes are the {\it only}
real physical candidates for which both the black-hole event hoizon and 
and the analogue sonic horizon may co-exist. Hence,
our application of the analogue Hawking effect to the theory of 
transonic
astrophysical accretion may be useful to compare
the properties of these two types of horizons.

However, in our work, the analogy has been applied to
describe the classical perturbation of the fluid in terms of a
field satisfying the wave equation in an effective geometry.
It is not our aim to provide a  formulation by which
the phonon field generated in this system could be quantized.
To
accomplish this task, one would need to show that the effective action for the
acoustic perturbation is equivalent to a field theoretical action
in curved space, and the corresponding commutation and dispersion
relations  should
directly follow (see, e.g. \cite{stuz}).
Such considerations are beyond the scope of
this paper.
Also note that for all  types of
accretion discussed here, the analogue temperature $T_{\rm AH}$
is many orders of magnitude  lower compared with the fluid temperature
of accreting matter.

In constructing a model for the disc height we have followed Abramowicz et al.\ \cite{discheight}, although
 a number of  other models  for the disc height exist in the literature
\cite{nt73,riffert,pariev,peitz,lasota}. The use of any other disc 
height model would not alter our  conclusion 
that black-hole accretion disc solutions form an important class 
of analogue gravity models. However, the numerical values of 
$T_{AH}$ and other related quantities would  be different for 
different disc heights.

In this work, the viscous transport of the angular momentum is not 
explicitly taken into account. Viscosity, however, is quite a subtle 
issue in studying the analogue effects for disc accretion.
Thirty two years after the discovery of
standard accretion disc theory \cite{nt73,ss73}, exact modeling of viscous 
transonic black-hole accretion, including
proper heating and cooling mechanisms, is still quite an arduous task, even for a 
Newtonian flow.
 On the other hand,
from the analogue model point of view, viscosity
is likely to destroy  Lorenz invariance, and hence the assumptions behind building up an
analogue model may not be quite  consistent.
Nevertheless, extremely large radial velocity
close to the black hole implies $\tau_{inf}\ll \tau_{visc}$, where $\tau_{inf}$ and
$\tau_{visc}$ are the infall and the viscous time scales, respectively. 
Large radial velocities even at larger distances are due to the fact
that the angular momentum content of the accreting fluid
is relatively low \cite{belo,igubelo,proga}.
 Hence,
our assumption of inviscid flow is not unjustified from 
an astrophysical point of view.
However, 
one of the most significant effects of the introduction of viscosity 
would be the reduction of the angular momentum.
We found that the location of the sonic points
anti-correlates with $\lambda$, i.e. weakly rotating flow makes the
dynamical velocity gradient steeper, which indicates that for
viscous flow the acoustic horizons will be pushed further out and the flow would
become supersonic at a larger distance for the same set of other initial
boundary conditions.

In our model, we have performed the computation of $T_{AH}$ for 
non-rotating black holes. Recently, the spacetime geometry on the
equatorial slice through a Kerr black hole has been shown to be 
equivalent to the geometry experienced by phonons in a rotating 
fluid vortex \cite{silke}. Since many astrophysical 
black holes are expected to possess non-zero spin (the Kerr
parameter $a$), a clear understanding of the influence of spin on 
analogue models will be of great importance.  Our initial calculation 
indicates that the black-hole spin {\it enhances} the analogue 
effect, i.e. the analogue Hawking temperature  
increases with the Kerr parameter $a$.

In connection to the acoustic geometry, one can define 
an `anti-trapped surface' to be a hypersurface in which 
the fluid flow will be outward directed with the normal component 
of the three-velocity greater than the local speed of sound. In stationary geometry, an anti-trapped surface will 
naturally be constructed by the collection of sonic 
points corresponding to a spherically symmetric or
axisymmetric transonic wind solution emanating out from an astrophysical
source. 
Transonic outflow (wind) is ubiquitous in astrophysics,
spanning a wide range from solar/stellar winds to  large-scale 
outflows from active galaxies, quasars, galactic micro-quasars
and energetic gamma ray bursts (GRB). In section \ref{sec:gcs}, we have shown how
to identify the critical and the sonic points corresponding to the
wind solutions. Such a formalism can be useful in studying
the transonic properties of outflow from astrophysical sources.
Hence our formalism presented in this paper can be applied to study
the analogue effects in transonic winds as well. Recently Kinoshita et al.\,
\cite{kinoshita} performed the causality analysis of the
spherical GRB outflow using the concept of effective acoustic geometry.
Such an investigation can be extended into a more robust form by 
incorporating our work to study the causal structure of the transonic GRB 
outflows in axisymmetry, i.e. for energetic directed outflow
originating from a black-hole accretion disc system progenitor.

In recent years, considerable attention has been focused on the study of 
gravitational collapse of massive matter clump, in particular, 
on
the investigation of the 
final fate of such collapse (for a review see, e.g. \cite{krolak}).
Goswami and Joshi \cite{panu} have studied  the role of the equation of
state and initial data in determining the final fate of the continual spherical 
collapse of barotropic fluid 
in terms of naked singularities and the black-hole formation.
It is  tempting to study the analogue effects in such a collapse
model. Since at some stage the  velocity of the collapsing 
fluid will exceed the velocity of local acoustic perturbation 
one might encounter a
sonic horizons at the radial locations of the
corresponding transonic points in a stationary configuration.
 One should, however, be careful about the issue that 
 many results in
analogue models are based on the assumption of a stationary flow,
whereas a  collapse scenario is a full time dependent  dynamical 
process. 

The correspondence between general relativity and analogue gravity has
 so far been exploited only on a kinematical, i.e.
geometrical level. The analogue gravity systems lack a proper dynamical
scheme, such as Einstein's field equations in general relativity and
hence the analogy is not
complete. 
A certain progress in this direction has recently been made 
by Cadoni and Mignemi
\cite{cadoni,mignemi} who have established a dynamical correspondence between
analogue and dilaton gravity in 1+1 dimensions. 
We believe that our approach in which
an arbitrary background geometry serves as a source for
 fluid dynamics may shed a new light towards a full analogy between  
general relativity and analogue gravity.

\appendix

\section{Derivation of the acoustic wave equation}
\label{appendix}
\noindent
This derivation is based in part on references
\cite{landau,visser}.
The continuity and Euler's equations may be
expressed as:
\begin{equation}
\frac{\partial{\rho}}{\partial{t}}+{\nabla}\cdot \left(\rho{\vec{u}}\right)
\end{equation}
\begin{equation}
\rho\frac{d{\vec {u}}}{dt}{\equiv}
\rho\left[\frac{\partial{\vec {u}}}{\partial{t}}
+\left({\vec {u}}\cdot {\nabla}\right){\vec {u}}\right]
=-\nabla{p}+{\vec {F}}
\label{eq002}
\end{equation}
with ${\vec {F}}$ being the sum of all external forces acting on the 
fluid wich may be expressed in terms of a potential
\begin{equation}
{\vec {F}}=-{\rho}{\nabla}{\Phi},
\end{equation}
Euler's equation may now be recast in the form
\begin{equation}
\frac{\partial{\vec {u}}}{\partial{t}}=
{\vec {u}} \times ( \nabla \times {\vec {u}})  - {1\over\rho} \nabla p
- \nabla\left( {{1\over2}} u^2 + \Phi\right)
\label{eq003}
\end{equation}
Next we assume the fluid to be inviscid, irrotational, and
barotropic. Introducing the specific enthalpy $h$ such that
\begin{equation}
{\nabla}h=\frac{\nabla{p}}{\rho}
\label{eq004}
\end{equation}
and  the velocity potential $\psi$ for which 
${\vec {u}}=-{\nabla}\psi$, 
Eq. (\ref{eq003}) may be writen as
\begin{equation}
-\frac{\partial{\psi}}{\partial{t}} + h + {1\over2} (\nabla\psi)^2
 + \Phi = 0
\end{equation}

One now linearizes the continuity and  Euler's equation around some 
unperturbed background flow variables $\rho_0$, $p_0$, $\psi_0$.
Introducing
\begin{eqnarray}
\rho=\rho_0+\epsilon \rho_1 +{\cal O}(\epsilon^2), & & p=p_0+\epsilon p_1 +{\cal O}(\epsilon^2) ,
\nonumber \\
\psi=\psi_0+\epsilon \psi_1 +{\cal O}(\epsilon^2),
 & & h=h_0+\epsilon h_1,
\label{eq006}
\end{eqnarray} 
from the continuity equation we obtain
\begin{equation}
\frac{\partial\rho_0}{\partial{t}}+
\nabla\cdot(\rho_0 \; {\vec {u}}_0) = 0; \;\;\;\;\;
\frac{\partial\rho_1}{\partial{t}}+
\nabla\cdot(\rho_1 \; {\vec {u}}_0 + \rho_0 \; {\vec {u}}_1) = 0.
\end{equation}
Equation (\ref{eq004}) implies
\begin{equation}
h_1=p_1\frac{dh}{dp}=
\frac{p_1}{\rho_0}\, .
\label{eq007}
\end{equation}
Using this the linearized Euler equation reads
\begin{equation}
-\frac{\partial{\psi_0}}{\partial{t}} + h_0 + {1\over2} (\nabla\psi_0)^2
 + \Phi = 0; \;\;\;\;\;
-\frac{\partial{\psi_0}}{\partial{t}}
+ {p_1\over\rho_1} - {\vec {u}}_0 \cdot \nabla\psi_1 = 0.
\end{equation}
Re-arrangement of the last equation 
together  with the barotropic assumption yields
\begin{equation}
\rho_1 =
{\partial \rho\over\partial p} \; p_1 =
{\partial \rho\over\partial p} \; \rho_0  \;
( \partial_t \psi_1 + {\vec {u}}_0 \cdot \nabla\psi_1 )\, .
\end{equation}
Substitution of this  into the linearized continuity equation 
gives the sound wave equation 
\begin{equation}
-\frac{\partial}{\partial{t}}
\left[ {\partial\rho\over\partial p} \; \rho_0 \;
            \left(\frac{\partial\psi_1}
{\partial{t}} + {\vec {u}}_0 \cdot \nabla\psi_1\right)
     \right]+
\nabla \cdot
     \left[ \rho_0 \; \nabla\psi_1
            - {\partial\rho\over\partial p} \; \rho_0 \; {\vec {v}}_0 \;
              \left( \frac{\partial\psi_1}{\partial{t}} 
+ {\vec {v}}_0 \cdot \nabla\psi_1\right)
     \right]
=0 .
\label{eq008}
\end{equation}
Next, we define the local speed of sound by 
\begin{equation}
c_s^2={\partial{p}}/{\partial\rho},
\end{equation}
where the partial derivative is taken at constant specific entropy.
With help of the $4\times{4}$ matrix 
\begin{equation}
f^{\mu\nu} \equiv
{\rho^0}
\left[ \matrix{-{I}&\vdots&-{\vec {u}}\cr
               \cdots\cdots&\cdot&\cdots\cdots\cdots\cdots\cr
               -{\vec {u}}&\vdots&( c_s^2 - {{u}^2} )\cr }
\right]
\label{eq009}
\end{equation}
where ${I}$ is the $3\times3$ identity matrix,
 one can put  Eq. (\ref{eq008}) to the form 
 \begin{equation}
{\partial}_{\mu}\left(f^{\mu{\nu}}{\partial}_{\nu}{\psi_1}\right)=0.
\end{equation}

\section*{Acknowledgements}
\noindent
TKD acknowledges the hospitality (in the form of 
visiting faculty position)
of the Department of Astronomy \& Astrophysics, TIFR.
The research of SD is partly supported by
the Kanwal Rekhi scholarship of the TIFR endowment fund.
SD also acknowledges the hospitality of the HRI during 
his visit.

{}

\end{document}